\documentclass{article}
\usepackage[english]{babel}

\usepackage{amsmath}
\usepackage{amssymb}

\usepackage{graphicx}
\usepackage[colorlinks=true,linkcolor=blue, citecolor=cyan, urlcolor=cyan]{hyperref}      

\usepackage{graphicx}
\usepackage{caption}
\usepackage{subcaption}
\usepackage{chngcntr}
\usepackage{caption}

\begin{document}

\title{{\Large Competitive division of a mixed manna\thanks{%
Support from the Basic Research Program of the National Research University
Higher School of Economics is gratefully acknowledged. Moulin thanks the
Simmons institute for Theoretical Computing for its hospitality during the
Fall 2015.Sandomirskiy is partially supported by the grant 16-01-00269 of
the Russian Foundation for Basic Research. The comments of William Thomson,
three anonymous referees and the Editor have been especially useful. } 
\thanks{%
This paper subsumes \cite{BMSY} and \cite{BMSY1}.}}}
\author{{\large Anna Bogomolnaia}$^{\bigstar \spadesuit }${\large , Herv\'{e}
Moulin}$^{\bigstar \spadesuit }${\large ,} \and {\large Fedor Sandomirskiy}$%
^{\spadesuit } ${\large , and Elena Yanovskaya}$^{\spadesuit }${\large .}
}
\date{{\large $^{\bigstar } $ \textit{University of Glasgow}\\
\vskip 0.1cm 
$^{\spadesuit }$ 
\textit{Higher School of Economics, St Petersburg}}}
\maketitle

\begin{abstract}
A mixed manna\ contains \textit{goods }(that everyone likes), \textit{bads}
(that everyone dislikes), as well as items that are \textit{goods} to some
agents, but \textit{bads} or satiated to others.

If all items are goods and utility functions are \textit{homothetic},
concave (and monotone), the \textit{Competitive Equilibrium with Equal
Incomes} maximizes the Nash product of utilities: hence it is \textit{%
welfarist} (determined utility-wise by the feasible set of profiles),
single-valued and easy to compute.

We generalize the Gale-Eisenberg Theorem to a mixed manna. The Competitive
division is still welfarist and related to the product of utilities or
disutilities. If the zero utility profile (before any manna) is Pareto
dominated, the competitive profile is unique and still maximizes the product
of utilities. If the zero profile is unfeasible, the competitive profiles
are the critical points of the product of \textit{dis}utilities on the
efficiency frontier, and multiplicity is pervasive. In particular the task
of dividing a mixed manna is either good news for everyone, or bad news for
everyone.

We refine our results in the practically important case of linear
preferences, where the axiomatic comparison between the division of goods
and that of bads is especially sharp. When we divide goods and the manna
improves, everyone weakly benefits under the competitive rule; but no
reasonable rule to divide bads can be similarly \textit{Resource Monotonic}.
Also, the much larger set of Non Envious and Efficient divisions of bads can
be disconnected so that it will admit no continuous selection.
\end{abstract}

\section{Introduction and main result}

The literature on fair division of private commodities, with few exceptions
discussed in Section 3, focuses almost exclusively on the distribution of
disposable commodities, i. e., desirable \textit{goods} like a cake (\cite%
{St}), family heirlooms (\cite{PZ}), the assets of divorcing partners (\cite%
{BT}), office space between co-workers, seats in overdemanded business
school courses (\cite{SU}, \cite{BC}), computing resources in peer-to-peer
platforms (\cite{GZHKSS}), and so on. Obviously many important fair division
problems involve \textit{bads} (non disposable items generating disutility):
family members distribute house chores, workers divide job shifts (\cite{B})
like teaching loads, cities divide noxious facilities, managers allocate
cuts within the firm, and so on. Moreover the bundle we must divide (the 
\textit{manna}) often contains the two types of items: dissolving a
partnership involves distributing its assets as well as its liabilities,
some teachers relish certain classes that others loathe, the land to be
divided may include polluted as well as desirable areas, and so on. And the
manna may contain items, such as shares in risky assets, or hours of
baby-sitting, over which preferences are single-peaked without being
monotone, so they will not qualify as either \textquotedblleft
good\textquotedblright\ or \textquotedblleft bad\textquotedblright , they
are \textquotedblleft satiable\textquotedblright\ items. Of course each item
may be a good to some agents, a bad to others, and satiable to yet other
agents. We speak in this case of dividing a \textit{mixed manna}.

Although the fair division literature pays some attention to the case of a
\textquotedblleft bad\textquotedblright\ manna, our paper is, to the best of
our knowledge, the first to address the case of a mixed manna.

To see why it is genuinely more complicated to divide a mixed rather than a
good or a bad manna, consider the popular fairness test of \textit{%
Egalitarian Equivalence} (EE) due to Pazner and Schmeidler (\cite{PS}). A
division of the manna is EE if everyone is indifferent between her share and
some common reference share: with mixed items this property may well be
incompatible with Efficiency.\footnote{%
Two agents $1,2$ share (one unit of) two items $a,b$, and their utilities
are linear: $u_{1}(z_{1})=z_{1a}-2z_{1b}$; $u_{2}(z_{2})=-2z_{2a}+z_{2b}$.
The only efficent allocation gives $a$ to $1$ and $b$ to $2$. In an EE
allocation $(z_{1},z_{2})$ there is some $y\geq 0$ such that $%
u_{i}(z_{i})=u_{i}(y)$ for $i=1,2$. This implies $%
u_{1}(z_{1})+u_{2}(z_{2})=-(y_{a}+y_{b})$ so that $z$ is not efficient.} The
news is much better for the division proposed by microeconomists four
decades ago (\cite{V}), the \textit{Competitive Equilibrium with Equal
Incomes }(here competitive division, for short). Existence is guaranteed
when preferences are convex, continuous, but not necessarily monotonic and
possibly satiated: see e. g., \cite{SS0}, \cite{MC}. And this division
retains the key normative properties of Efficiency, No Envy, and Core
stability from equal initial endowments (see Lemma 1 Section 4).

A striking result by Gale, Eisenberg, and others (\cite{Ga}, \cite{E}, \cite%
{C}, \cite{SS}) shows that in the subdomain of \textit{homothetic }(as well
as concave and continuous)\textit{\ }utilities the competitive division of 
\textit{goods} obtains by simply maximizing the product of individual
utilities. This is remarkable for three reasons. First the \textquotedblleft
resourcist\textquotedblright\ concept of competitive division guided by a
price balancing Walrasian demands, has an equivalent \textquotedblleft
welfarist\textquotedblright \textit{\ }interpretation as the Nash bargaining
solution of the feasible utility set. Second, the competitive utility
profile is unique because by the latter definition it solves a strictly
convex optimization program; it is also computationally easy to find and
continuous with respect to the parameters of individual utilities (\cite{Va}%
, \cite{JV}); all these properties fail under general Arrow-Debreu
preferences. Finally the result is broadly applicable because empirical work
relies mostly on homothetic utilities, that include additive, Cobb Douglas,
CES, Leontief, and their linear combinations. So the Gale Eisenberg theorem
is arguably the most compelling practical vindication of the competitive
approach to the fair division of goods.

We generalize this result to the division of a mixed manna under concave,
continuous and homothetic preferences. We show that the welfarist
interpretation of the competitive division is preserved: the set of feasible
utility profiles is still all we need to know to identify the competitive
utility profiles (those associated with a competitive division of the
items). On the other hand there may be many different such profiles, and in
that case they no longer solve a convex program: computational simplicity
and continuity as above are lost.

We also show that division problems are of three types, and that a very
simple welfarist property determines their type. Keep in mind that, by
homotheticity, the zero of utilities corresponds to the ex ante state of the
world without any manna to divide. Call an agent \textquotedblleft
attracted\textquotedblright\ if there is a share of the manna giving her
strictly positive utility, and \textquotedblleft repulsed\textquotedblright\
if there is none, that is to say zero is her preferred share.

If it is feasible to give a positive utility to all attracted agents, and
zero to all repulsed ones, we call this utility profile \textquotedblleft
positive\textquotedblright\ and speak of a \textquotedblleft
positive\textquotedblright\ problem. Then the competitive utility profile is
positive for attracted agents and maximizes the product of the attracted
agents' utilities over positive profiles; just like in Gale Eisenberg this
utility profile is unique and easy to compute. Also, the arrival of the
manna is (weakly) good news for everyone.

If on the other hand the efficiency frontier contains allocations where 
\textit{everyone} gets a strictly negative utility, we call the problem
\textquotedblleft negative\textquotedblright . Then the competitive utility
profiles are the \textit{critical points} (for instance local maxima or
minima) of the product of all \textbf{dis}utilities on the intersection of
the efficiency frontier with the (strictly) negative orthant:\footnote{%
See the precise definition in Section 5.} we may have multiple such
profiles, and we expect computational difficulties.\footnote{%
Selecting the competitive utility profiles \textit{maximizing} the product
of \textit{dis}utilities on this part of the efficiency frontier almost
surely gives a unique utility profile (Lemmas 3 and 4), but does not
eliminate the computational and continuity issues, as explained by
Proposition 3.} Moreover, the arrival of the manna is strictly bad news for
everyone.

Finally the \textquotedblleft null\textquotedblright\ problems are those
knife-edge cases where the zero utility profile is efficient: it is then the
unique competitive utility profile, and the arrival of the manna is no news.

\section{The case of linear preferences}

The simplest subdomain of the homothetic domain just discussed is that of
linear preferences, represented by additive utilities. Its practical
relevance is vindicated by user-friendly platforms like SPLIDDIT or ADJUSTED
WINNER\footnote{%
\url{www.spliddit.org/}; \url{www.nyu.edu/projects/adjustedwinner/}}, computing fair
outcomes in a variety of problems including the division of manna. Visitors
of these sites must distribute 100 points over the different items, and
these \textquotedblleft bids\textquotedblright\ are interpreted as fixed
marginal utilities, positive for goods, negative for bads, and zero for a
satiated item.\ At the cost of ignoring complementarities between items,
this makes the report of preferences fairly easy, eschewing the complex task
of reporting full fledged preferences when we have more than a handful of
items.\footnote{%
Similarly practical combinatorial auctions never ask buyers to report a
ranking of all subsets of objects, (\cite{BL}, \cite{VV}, \cite{CSS}).} The
proof of the pudding is in the eating: tens of thousands of visitors have
used these sites since 2014, fully aware of the interpretation of their bids
(\cite{GP}).

If $N$ is the set of agents and $A$ that of items, a profile of additive
utilities is described by a $N\times A$ matrix $u=[u_{ia}]$ with $i\in
N,a\in A$; agent $i$'s utility for allocation $z_{i}\in 
%TCIMACRO{\U{211d} }%
%BeginExpansion
\mathbb{R}
%EndExpansion
_{+}^{A}$ is $u_{i}(z_{i})=\sum_{A}u_{ia}z_{ia}$. If all items are \textit{%
goods}, the marginal utilities $u_{ia}$ are all non negative and in the
terminology just introduced the problem is \textit{positive}: the classic
Gale Eisenberg result applies and the competitive utility profile is the
unique Nash bargaining solution. If all items are \textit{bads}, the
marginal utilities $u_{ia}$ are all non positive and the problem is \textit{%
negative}.

In the additive domain we evaluate first the potentially severe multiplicity
of competitive divisions in negative problems, illustrated in the numerical
example below. Next we propose an invariance property in the spirit of
Maskin Monotonicity characterizing the competitive division rule for any
problem, positive, negative or null.

On the other hand we prove some strong impossibility results for all-bads
problems (hence for negative ones as well): they limit the appeal of any
division rule guaranteeing No Envy, or simply a fair share of the manna to
every participants. Therefore the contrast between positive and negative
problems goes beyond the competitive approach, which is somewhat
counter-intuitive: just like labor is time not spent on leisure, allocating $%
z_{ia}$ units of bad $a$ to $i$ is the same as exempting her from eating $%
\omega _{a}-z_{ia}$ units of $a$ (where $\omega _{a}$ is the amount of bad $%
a $ in the manna). But note that we must distribute $(|N|-1)\omega _{a}$
units of the $a$-exemption, while each agent can eat at most $\omega _{a}$
units of it: these additional capacity constraints create the normative
differences that we identify.

\paragraph{A numerical example}

\begin{figure}[h!]
\centering
\begin{subfigure}{0.48\linewidth}
\includegraphics[width=1\linewidth, clip=true, trim = 2.5cm 6.8cm 2.5cm 6.5cm]
%trim = 4.0cm 6.8cm 4.0cm 6.5cm]
{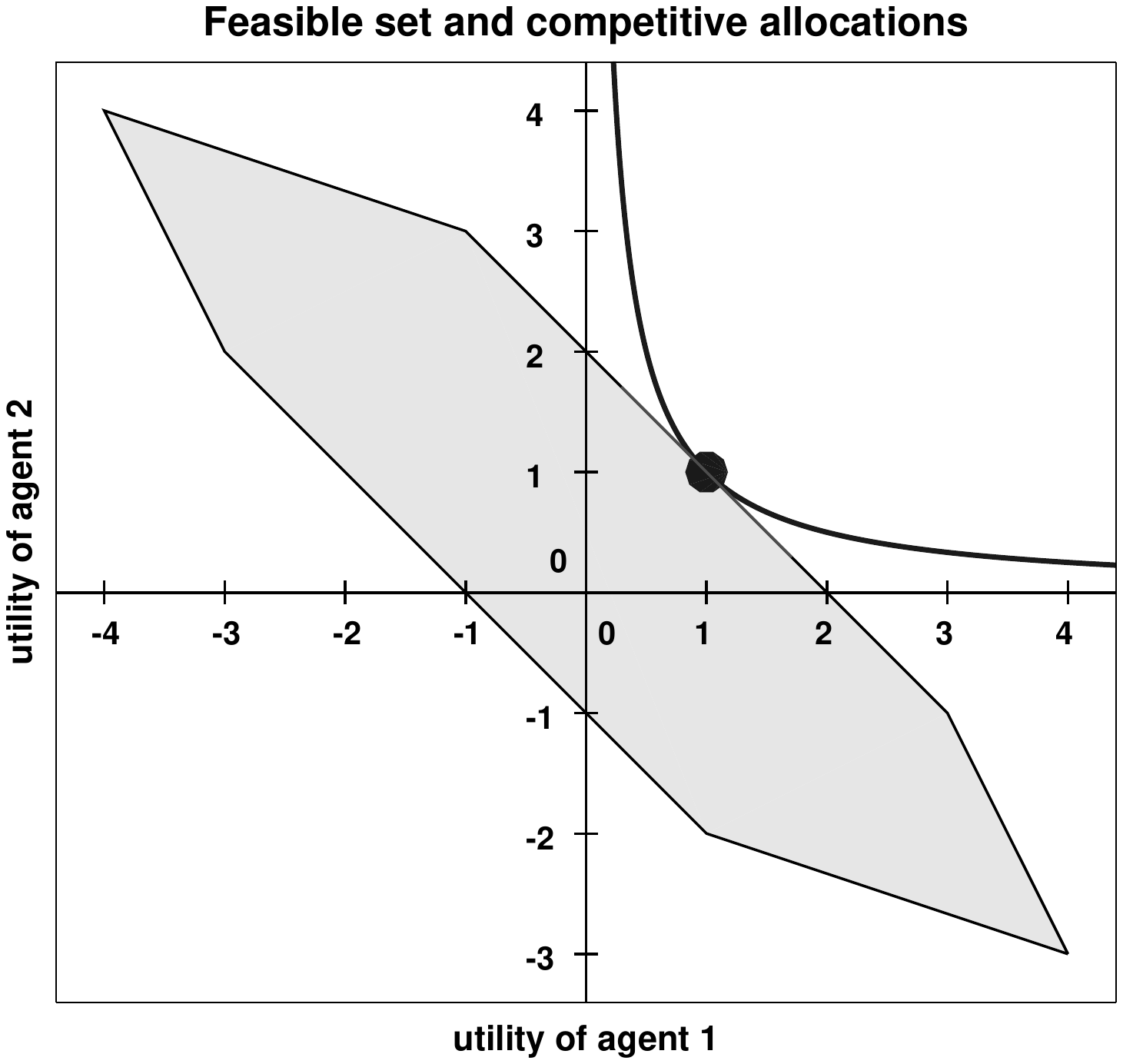}
\end{subfigure}
~
\begin{subfigure}{0.48\linewidth}
\includegraphics[width=1\linewidth, clip=true, trim = 2.5cm 6.8cm 2.5cm 6.5cm]
{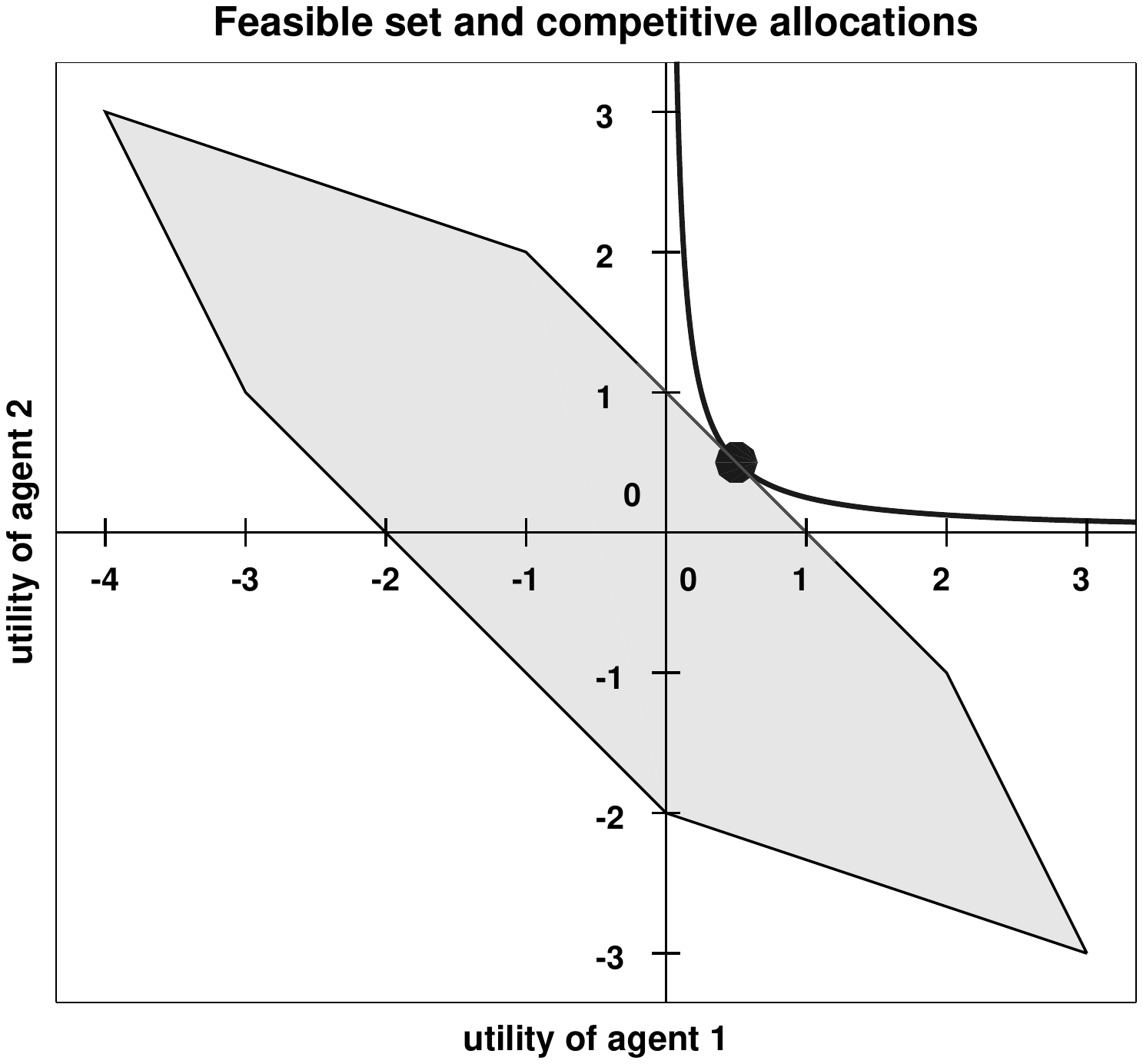}
\end{subfigure}
\caption*{Figure 1 ($\lambda=4,3$)}
\end{figure}

We start with a two agent, three items sequence of examples illustrating the
complicated pattern of competitive allocations in negative problems. We have
two agents $N=\{1,2\}$, three items $A=\{a,b,c\}$, one unit of each item,
and marginal utilities are%
\begin{equation*}
\begin{array}{cccc}
& a & b & c \\ 
u_{1} & -1 & -3 & \lambda \\ 
u_{2} & -2 & -1 & \lambda%
\end{array}%
\end{equation*}

\begin{figure}[h!]
\centering
\begin{subfigure}{0.48\linewidth}
\includegraphics[width=1\linewidth, clip=true, trim = 2.5cm 6.8cm 2.5cm 6.5cm]
{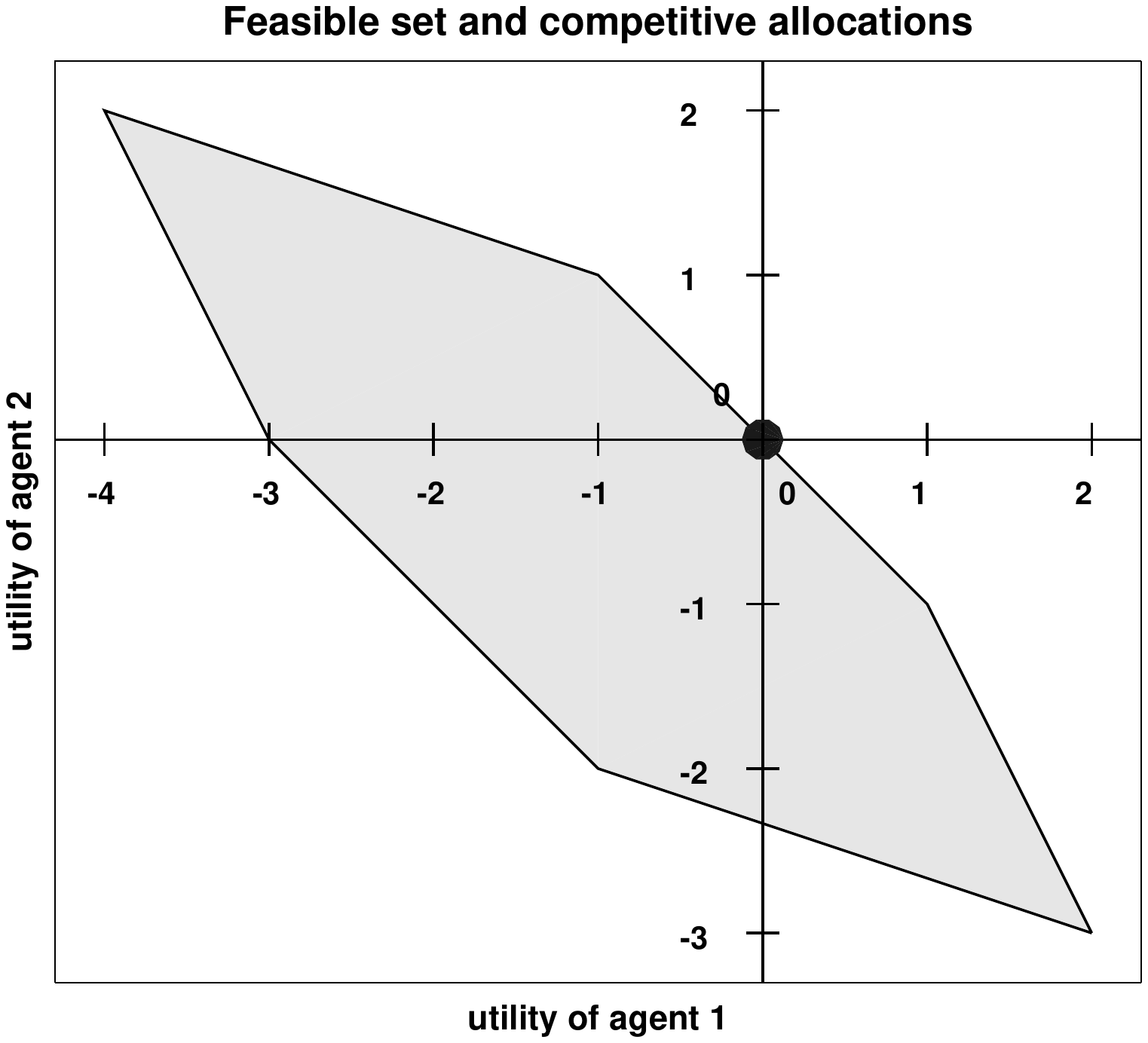}
\end{subfigure}
~
\begin{subfigure}{0.48\linewidth}
\includegraphics[width=1\linewidth, clip=true, trim = 2.5cm 6.8cm 2.5cm 6.5cm]
{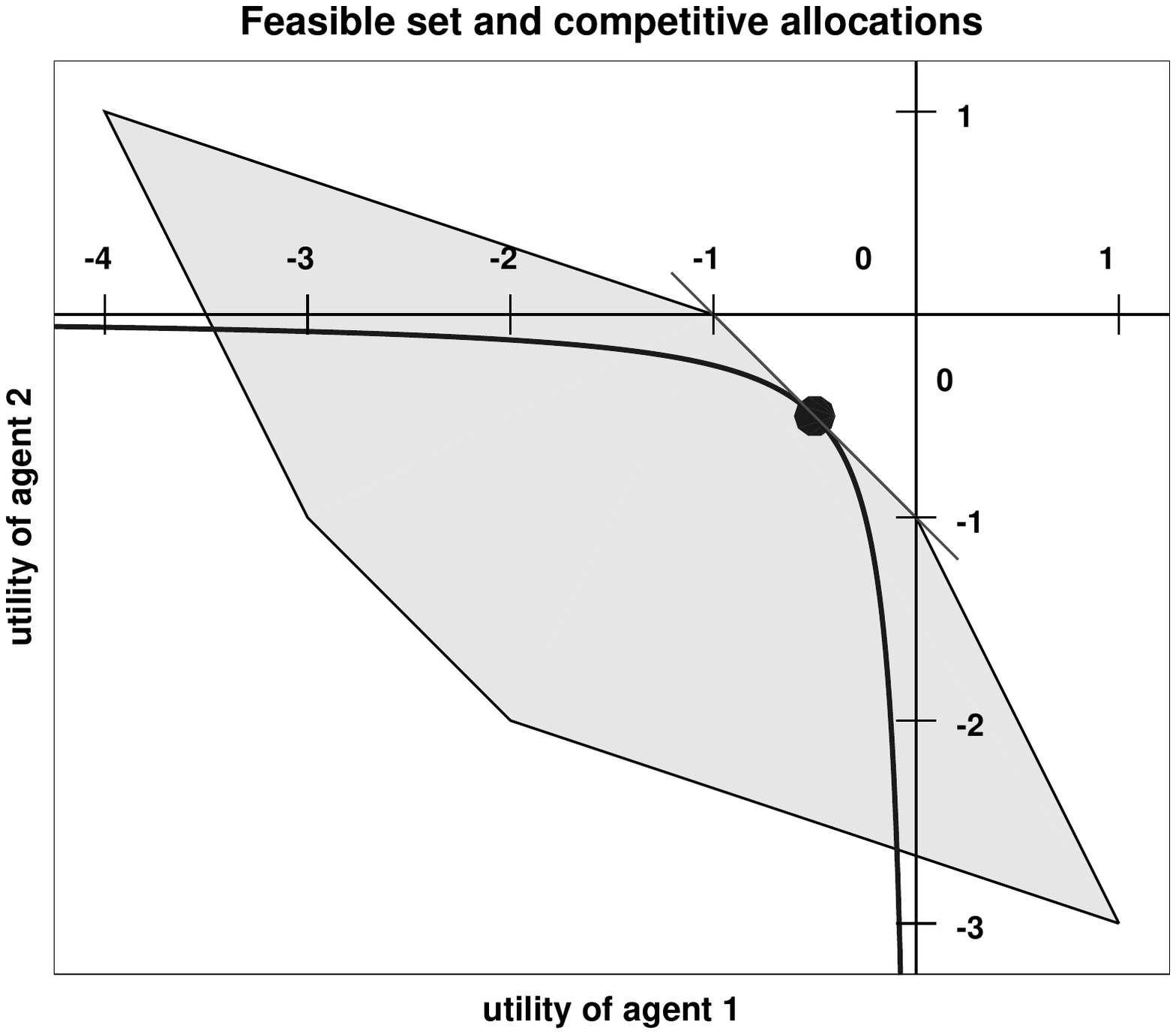}
\end{subfigure}
%\caption*{Figure 1 ($\lambda=2,1$)}
\end{figure}

\noindent Items $a,b$ are bads; as $\lambda $ takes all integer values from $%
4$ to $-3$, item $c$ goes from good to satiated ($\lambda =0$) to bad. For $%
\lambda =4,3$ the problem is positive; it is null for $\lambda =2$, then
negative from $\lambda =1$ to $-3$. Figure 1 shows in each case the set of
feasible utility profiles, and the competitive utility profiles. Their
number varies from $1$ to $4$. For instance if $\lambda =-1$ all items are
bads and the four competitive utility profiles are $(-1,-2)$, $(-1.5,-1.5)$, 
$(-2,-1)$, $(-2.5,-0.83)$. In Section 5 (Lemmas 3 and 4) we propose to
select the profile $(-1.5,-1.5)$ maximizing the product of \textit{dis}%
utilities; the corresponding allocation is $z_{1}=(1,0,\frac{1}{2})$, $%
z_{1}=(0,1,\frac{1}{2})$. Note that for $\lambda =1$ (Figure 1.e) this
maximum is achieved by the competitive allocation most favourable to agent $%
1 $.
\begin{figure}[h!]
\centering

\begin{subfigure}{0.48\linewidth}
\includegraphics[width=1\linewidth, clip=true, trim = 2.5cm 6.8cm 2.5cm 6.5cm]
{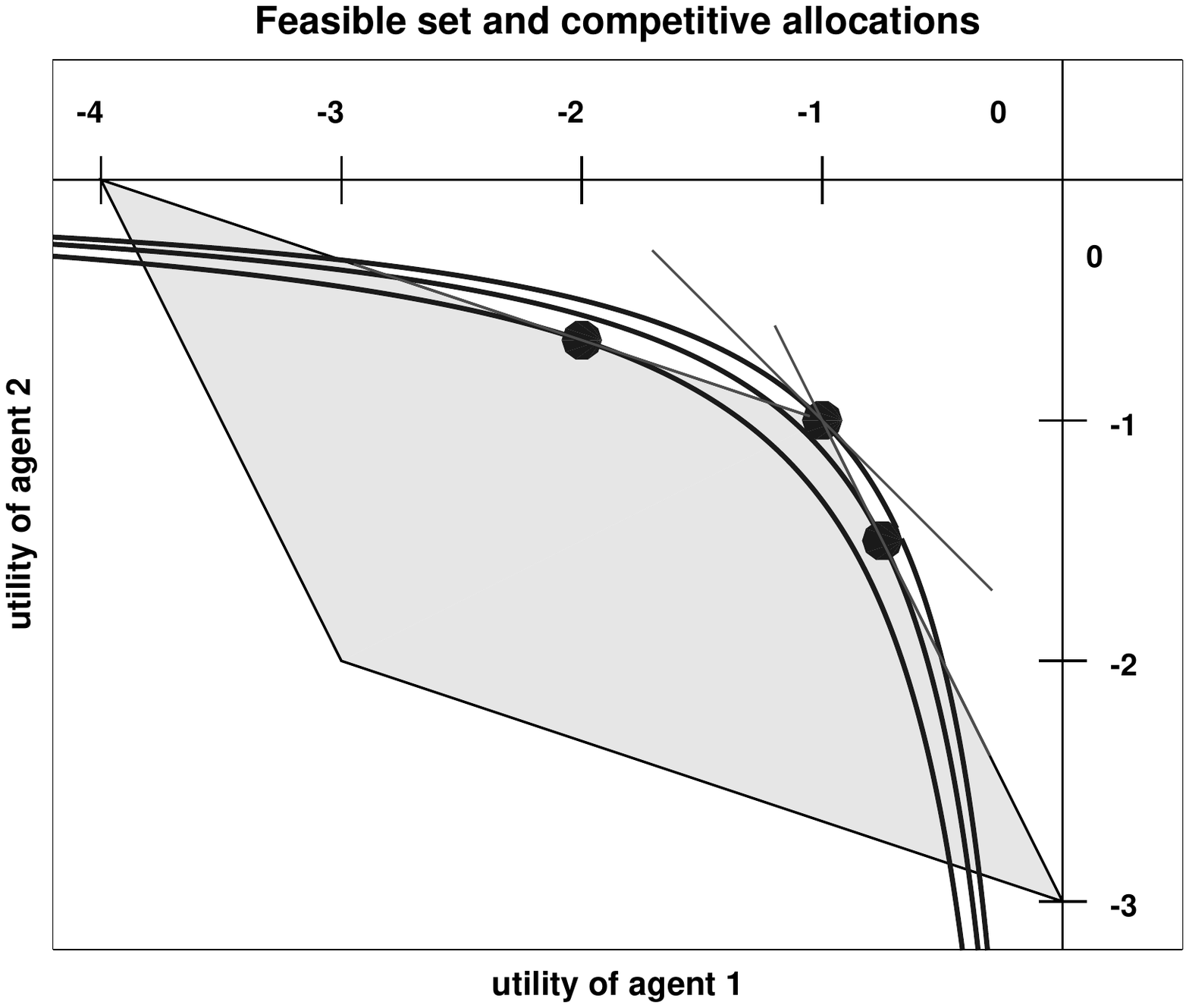}
\end{subfigure}
~
\begin{subfigure}{0.48\linewidth}
\includegraphics[width=1\linewidth, clip=true, trim = 2.5cm 6.8cm 2.5cm 6.5cm]
{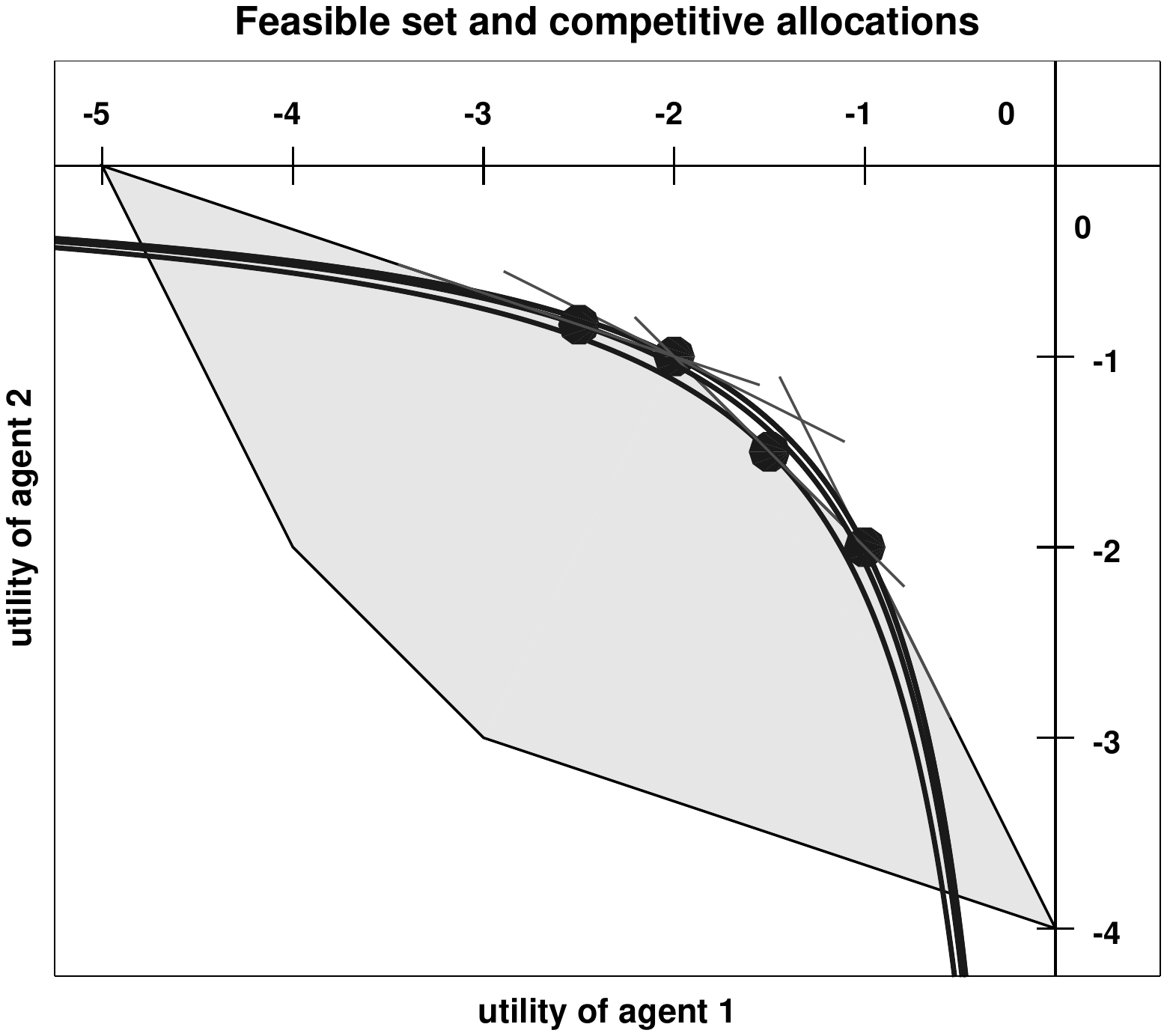}
\end{subfigure}

\caption*{Figure 1 ($\lambda=2,1,0,-1$)}
\end{figure}

In Subsection 6.1 we estimate the maximal number of welfare-wise different
competitive allocations in an \textit{all bads} (or negative) problem. This
number grows at least exponentially in the smallest of the number of agents
or bads (Proposition 1).

\begin{figure}[h!]
\centering
\begin{subfigure}[b]{0.48\linewidth}
\includegraphics[width=1\linewidth, clip=true, trim = 2.5cm 6.8cm 2.5cm 6.5cm]
{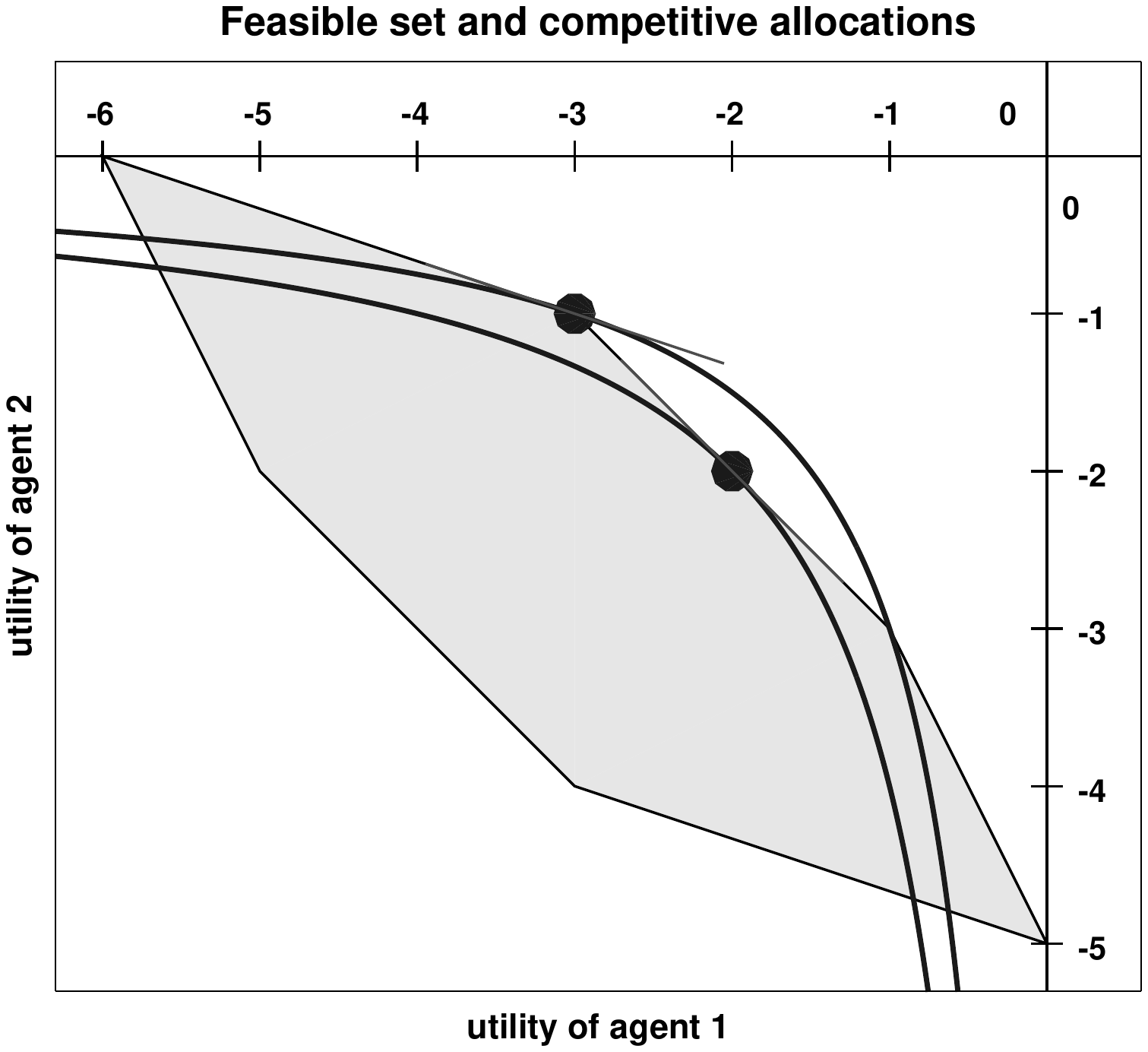}
\end{subfigure}
~
\begin{subfigure}[b]{0.48\linewidth}
\includegraphics[width=1\linewidth, clip=true, trim = 2.5cm 6.8cm 2.5cm 6.5cm]
{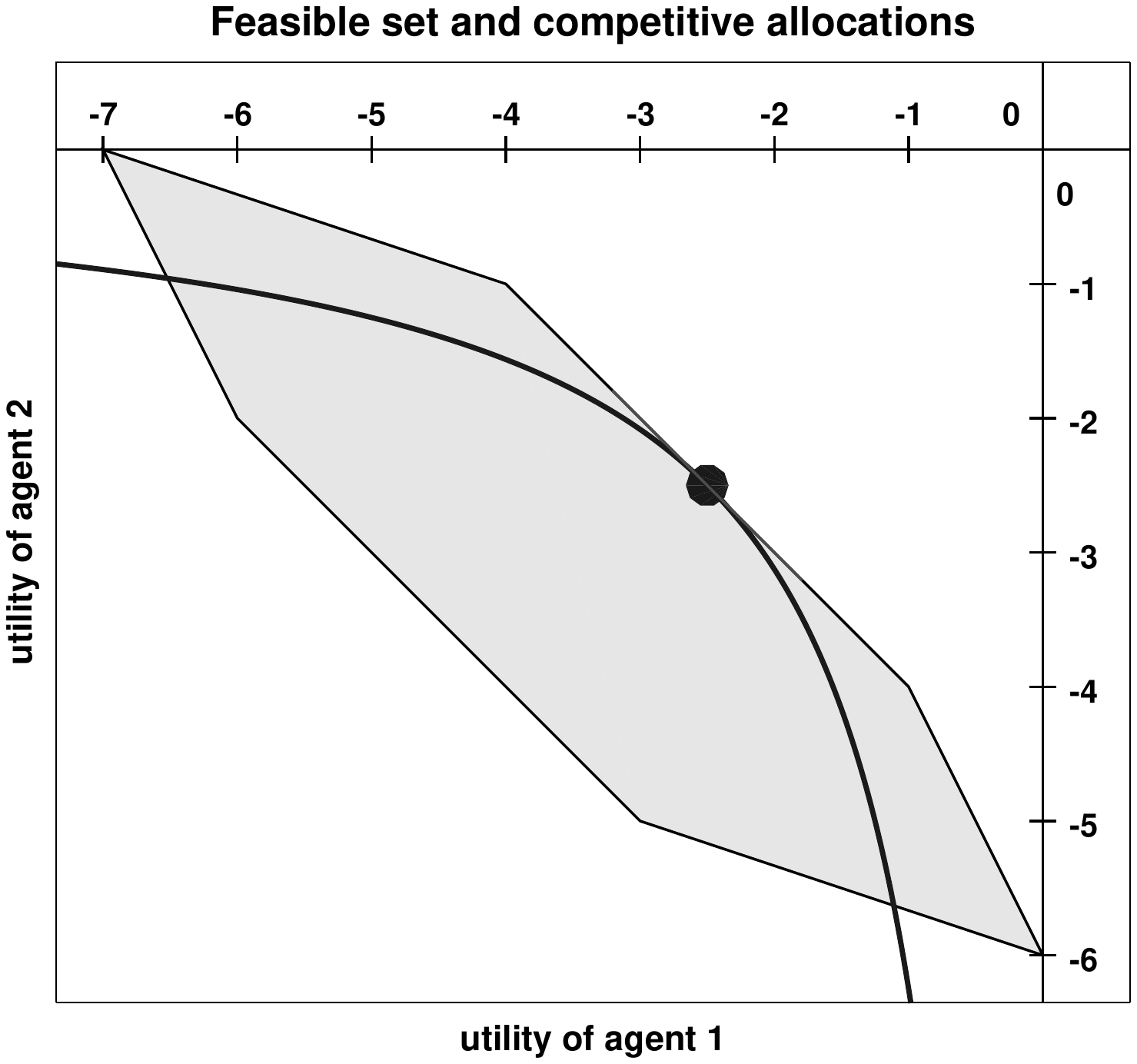}
\end{subfigure}
\caption*{Figure 1 ($\lambda=-2,-3$)}
\end{figure}

\paragraph{An axiomatic characterization of the competitive rule}

In the linear domain agents report marginal utilities, which we can
interpret as \textquotedblleft bids\textquotedblright\ for the different
items (as in \cite{SU}). Thus agent $i$'s bid $u_{ia}$ for item $a$ is
\textquotedblleft losing\textquotedblright\ if she ends up not consuming any 
$a$. \textit{Independence of Lost Bids\ }(ILB)\textit{\ }means that nothing
changes when we lower a losing bid: it remains losing and the allocation
selected by the rule does not change. The ILB axiom implies a weak incentive
property: misreporting on an item which I do not consume anyway (whether I
misreport or not) does not pay, and does not affect anyone else either. Here
is another consequence of ILB. Suppose item $a$ is a good for agent $1$, $%
u_{ia}>0$, but a bad for agent $2$, $u_{2a}<0$; by Efficiency agent $2$
consumes no $a$; then ILB says that the selected allocations do not change
if agent $2$'s bid for $a$ was zero instead of $u_{2a}$. In turn this means
that if an item is strictly good for someone, we can assume that it is
either good or satiated\ for everyone else.

The ILB axiom is a weak form of Maskin Monotonicity as explained in
Subsection 7.7. In combination with the requirement that all agents end up
on the same side of their zero utility, it promptly characterizes the
competitive rule for all mixed manna problems (Proposition~2).

\paragraph{Continuity and Monotonicity properties}

For a general problem with goods and bads, the set of competitive utility
profiles is an upper-hemi-continuous correspondence in the matrix of
marginal utilities. For positive problems it is single-valued, hence
continuous, but for negative problems it does not admit a continuous
single-valued selection. Proposition 3 strengthens this statement by
weakening the competitiveness requirement to the much less demanding test of
No Envy. In an \textit{all bads }problem with three or more agents, there is
no continuous single valued selection of the set of efficient and Non
Envious allocations; in particular with $n$ agents and two bads the
corresponding set of utility profiles can have up to roughly $\frac{2}{3}n$
connected components.

Our last result is also a (simple) impossibility statement. We use the
familiar axiom \textit{Resource Monotonicity} (RM) to draw another wedge
between positive and negative problems. RM is a solidarity requirement when
the manna improves: if we increase the amount of a unanimous good (an item
everyone likes), or decrease that of a unanimous bad, everyone should
benefit at least weakly.\footnote{%
RM has been applied to many other resource allocation problems with
production and/or indivisibilities. See the recent survey \cite{T1}.} In a
positive problem the (single-valued) competitive rule is Resource Monotonic,
but in an all bads problems (hence in negative problems as well), no
single-valued rule guaranteeing his \textit{Fair Share} to every agent%
\footnote{%
That is, no one is worse off than by consuming a $\frac{1}{n}$-th share of
every item. It is an uncontroversial fairness requirement, much weaker than
No Envy in the linear domain.} is Resource Monotonic (Proposition 4).

\paragraph{Contents}

After reviewing the literature (Section 3) and defining the model (Section
4), Section 5 states our generalization of Gale Eisenberg to mixed manna. We
focus in Section 6 on the subdomain of linear preferences and the four
propositions just described. All substantial proofs are in Section 7.

\section{Related literature}

\indent 1. Steinhaus' 1948 \textquotedblleft cake-division\textquotedblright\ model (%
\cite{St}), assumes linear preferences represented by atomless measures
over, typically, a compact euclidean set. It contains our model for goods as
the special case where the measures have piecewise constant densities.
Sziklai and Segal-Halevi (\cite{SS1}) show that it preserves the equivalence
of the competitive rule and the Nash product maximizer, and that this rule
is Resource Monotonic. The cake division literature pays some attention to
the division of a \textit{bad }cake, to prove the existence of envy-free
divisions of the cake (\cite{Su}, \cite{AS}), or to examine how the classic
algorithms by cuts and queries can or cannot be adapted to this case (\cite%
{BT}, \cite{RW}). It does not discuss the competitive rule for a bad cake.

2. The recent work in computational social choice discusses extensively the
fair division of goods (see the survey \cite{COMSOC}), recognizing the
practical convenience of additive utilities and the conceptual advantages of
the competitive solution in that domain (see \cite{M1}, \cite{Va}). For
instance Megiddo and Vazirani (\cite{MV}) show that the competitive utility
profile depends continuously upon the rates of substitution and the total
endowment; Jain and Vazirani (\cite{JV}) that it can be computed in time
polynomial in the dimension $n+m$ of the problem (number of agents and of
goods).

3. The fair division of \textit{indivisible} \textit{goods }with additive
utilities is a much studied variant of the standard model. The maximization
of the Nash product loses its competitive interpretation and becomes hard to
compute (\cite{L}), however it is envy-free \textquotedblleft up to at most
one object\textquotedblright\ (\cite{CKMPSW}) and can be efficiently
approximated for many utility domains (\cite{CG}, \cite{AGSS}, \cite{AMGV}, 
\cite{CDGJMV}). Also Budish (\cite{B}) approximates the competitive
allocation in problems with a large number of copies of several good-types
by allowing some flexibility in the number of available copies.

4. Our Proposition 2 is closely related to several axiomatic
characterizations of the competitive rule for the fair division of private 
\textit{goods}, in the much larger domain of Arrow-Debreu preferences. The
earliest results by Hurwicz\textbf{\ }(\cite{H}) and Gevers (\cite{G}) are
refined by Thomson\textbf{\ }(\cite{T2}) and Nagahisa (\cite{N}): any
efficient and Pareto indifferent rule meeting (some variants of) Maskin
Monotonicity (MM) must contain the competitive rule.\footnote{%
Another, logically unrelated characterization combines Consistency and
Replication Invariance (\cite{T3}) or Consistency and Converse Consistency (%
\cite{NPT}).} Our Independence of Lost Bids is weaker than MM in the linear
domain, so our Proposition 2 is a variant of these results in the case of
mixed items (and homothetic preferences).

5. The probabilistic assignment of goods with von Neuman Morgenstern
utilities is another fair division problem with linear and possibly satiated
preferences where Hylland and Zeckhauser (\cite{HZ}) and the subsequent
literature recommend (a version of) the competitive rule: e. g., \cite{HMPY}%
. That rule is no longer related to the maximization of the product of
utilities.

6. The purely welfarist axiomatic discussion of non convex bargaining
problems identifies the set of critical points of the Nash product among
efficient utility profiles as a natural generalisation of the Nash solution: 
\cite{Her}, \cite{SeSh}. This solution stands out also in the rationing
model of \cite{MM} where we divide utility losses instead of gains. The
latter is closer in spirit to our results for the division of bads.

\section{The model}

The set of agents is $N$, that of items is $A$; both are finite. The domain $%
\mathcal{H}(A)$ consists of all preferences on $%
%TCIMACRO{\U{211d} }%
%BeginExpansion
\mathbb{R}
%EndExpansion
_{+}^{A}$ represented by a real-valued utility function $v$ on $%
%TCIMACRO{\U{211d} }%
%BeginExpansion
\mathbb{R}
%EndExpansion
_{+}^{A}$ that is concave, continuous, and $1$-homothetic: $v(\lambda
y)=\lambda v(y)$ for all $\lambda \geq 0,y\in 
%TCIMACRO{\U{211d} }%
%BeginExpansion
\mathbb{R}
%EndExpansion
_{+}^{A}$. It is easily checked that if two such utility functions represent
the same preference, they differ by a positive multiplicative constant. All
our definitions and results are purely ordinal, i. e., independent of the
choice of the utility representations; we abuse language by speaking of
\textquotedblleft the utility function $v$ in $\mathcal{H}(A)$%
\textquotedblright .

The graph of a concave and continuous function $v$ on $%
%TCIMACRO{\U{211d} }%
%BeginExpansion
\mathbb{R}
%EndExpansion
_{+}^{A}$ is the envelope of its supporting hyperplanes, therefore it takes
the form $v(y)=\min_{k\in K}\{\alpha _{k}\cdot y+\beta _{k}\}$ for some $%
\alpha _{k}\in 
%TCIMACRO{\U{211d} }%
%BeginExpansion
\mathbb{R}
%EndExpansion
^{A},\beta _{k}\in 
%TCIMACRO{\U{211d} }%
%BeginExpansion
\mathbb{R}
%EndExpansion
$ and a possibly infinite set $K$. It is easy to see that $v$ is also
homothetic if and only if we can choose $\beta _{k}=0$ for all $k$. So the
simplest examples are the additive utilities $v(y)=\alpha \cdot y$ and the
piecewise linear utilities like $v(y)=\min
\{y_{a}+y_{b},4y_{a}-y_{b},4y_{b}-y_{a}\}$ for $A=\{a,b\}$, of which the
indifference contours are represented on Figure 2. Note that this utility is
not globally satiated, but for fixed $y_{b}$ it is satiated at $y_{a}=y_{b}$%
. For a smooth example of a non monotonic function in $\mathcal{H}(A)$
consider for example $v(y)=y_{b}\ln \{\frac{y_{a}}{y_{b}}+\frac{1}{2}\}$,
represented in Figure 3.

\begin{figure}
\vskip -5cm
\hskip 1cm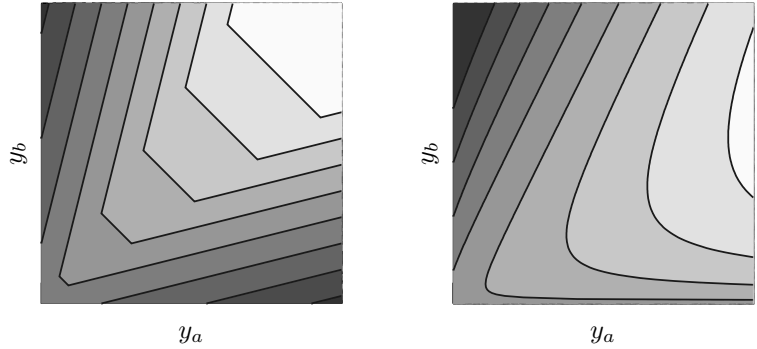
\vskip -0.3cm
\caption*{Figures 2 and 3}
\end{figure}

A \textit{fair division problem} is $\mathcal{P}=(N,A,u,\omega )$ where $%
u\in \mathcal{H}(A)^{N}$ is the profile of utility functions, and $\omega
\in 
%TCIMACRO{\U{211d} }%
%BeginExpansion
\mathbb{R}
%EndExpansion
_{+}^{A}$ is the manna; we assume $\omega _{a}>0$ for all $a$.

A \textit{feasible allocation} (or simply an \textit{allocation}) is $z\in 
%TCIMACRO{\U{211d} }%
%BeginExpansion
\mathbb{R}
%EndExpansion
_{+}^{N\times A}$ such that $\sum_{N}z_{ia}=\omega _{a}$ for all $a$, or in
a more compact notation $z_{N}=\omega $. The corresponding utility profile
is $U\in 
%TCIMACRO{\U{211d} }%
%BeginExpansion
\mathbb{R}
%EndExpansion
^{N}$ where $U_{i}=u_{i}(z_{i})$. Let $\mathcal{F}(N,A,\omega )$ be the set
of feasible allocations, and $\mathcal{U}(\mathcal{P)}$ the corresponding
set of utility profiles. We always omit $\mathcal{P}$ or $N,A$ if it creates
no confusion.

We call a feasible utility profile $U$ \textit{efficient }if it is not
Pareto dominated\footnote{%
That is $U\leq U^{\prime }$ and $U^{\prime }\in \mathcal{U}(\mathcal{%
P)\Longrightarrow }U^{\prime }=U$.}; a feasible allocation is efficient if
it implements an efficient utility profile.\smallskip

\noindent \textbf{Definition 1:} \textit{Given problem }$\mathcal{P}$\textit{%
\ a competitive division is a triple }$(z\in \mathcal{F},p\in 
%TCIMACRO{\U{211d} }%
%BeginExpansion
\mathbb{R}
%EndExpansion
^{A},\beta \in \{-1,0,+1\})$ \textit{where }$z$ \textit{is the competitive
allocation, }$p$\textit{\ is the competitive price and }$\beta $\textit{\
the individual budget. The allocation }$z$ \textit{is feasible and each }$%
z_{i}$\textit{\ maximizes }$i$\textit{'s utility in the budget set }$%
B(p,\beta )=\{y_{i}\in 
%TCIMACRO{\U{211d} }%
%BeginExpansion
\mathbb{R}
%EndExpansion
_{+}^{A}|p\cdot y_{i}\leq \beta \}$:\textit{\ }%
\begin{equation}
z_{i}\in d_{i}(p,\beta )=\arg \max_{y_{i}\in B(p,\beta )}\{u_{i}(y_{i})\}
\label{3}
\end{equation}%
\textit{Moreover }$z_{i}$\textit{\ minimizes }$i$\textit{'s wealth in her
demand set}%
\begin{equation}
z_{i}\in \arg \min_{y_{i}\in d_{i}(p,\beta )}\{p\cdot y_{i}\}  \label{12}
\end{equation}

\noindent \textit{We write} $CE(\mathcal{P)}$ \textit{for the set of
competitive allocations, and }$CU(\mathcal{P)}$ \textit{for the
corresponding set of utility profiles}.\smallskip

Existence of a competitive allocation can be derived from (much) earlier
results that do not require monotonic preferences (e.g., Theorem 1 in \cite%
{MC}; see also \cite{SS0}), but our main result in the next section gives
instead a constructive proof.

In addition to utility maximization (\ref{3}), property (\ref{12}) requires
demands to be \textit{parsimonious}: each agent spends as little as possible
for her competitive allocation. This requirement appears already in \cite{MC}%
: in its absence some satiated agents in $N_{-}$ may inefficiently eat some
items useless to themselves but useful to others.\footnote{%
For instance $N=\{1,2\},A=\{a,b\},\omega =(1,1)$ and $%
u_{1}(z_{1})=6z_{1a}+2z_{1b}$, $u_{2}(z_{2})=-z_{2b}$. The inefficient
allocation $z_{1}=(\frac{1}{3},1),z_{2}=(\frac{2}{3},0)$ meets (\ref{3}) for%
\textit{\ }$p=(\frac{3}{2},\frac{1}{2})$ and $\beta =1$. But $z_{2}^{\prime
}=(0,0)$ also gives zero utility to agent $2$ and costs zero, so $z_{2}$
fails (\ref{12}). The unique competitive division according Definition 1 is
efficient: $z_{1}=(1,1),z_{2}=0$, and $p=(\frac{1}{2},\frac{1}{2})$.}

Recall three standard normative properties of an allocation $z\in \mathcal{F}%
(N,A,\omega )$. It is \textit{Non Envious} iff $u_{i}(z_{i})\geq
u_{i}(z_{j}) $ for all $i,j$. It \textit{Guarantees} \textit{Fair Share}
utility \textit{iff} $u_{i}(z_{i})\geq u_{i}(\frac{1}{n}\omega )$ for all $i$%
. It is in the \textit{Weak Core from Equal Split} iff for all $S\subseteq N$
and all $y\in 
%TCIMACRO{\U{211d} }%
%BeginExpansion
\mathbb{R}
%EndExpansion
_{+}^{S\times A}$ such that $y_{S}=\frac{|S|}{n}\omega $, there is at least
one $i\in S$ such that $u_{i}(z_{i})\geq u_{i}(y_{i})$. When we divide goods
competitive allocations meet these three properties, even in the much larger
Arrow Debreu preference domain. This is still true with mixed
items.\smallskip

\noindent \textbf{Lemma 1 }\textit{A competitive allocation is efficient; it
is No Envious, Guarantees Fair Share, and is in the Weak Core from Equal
Split.}

\noindent \textbf{Proof}. No Envy is clear. Fair Share Guaranteed holds
because $B(p,\beta )$ contains $\frac{1}{n}\omega $. We check Efficiency. If 
$(z,p,\beta )$ is a competitive division and $z$ is Pareto-dominated by some 
$z^{\prime }\in \mathcal{F}$, then for all $i\in N$ we must have $%
(p,z_{i}^{\prime })\geq (p,z_{i})$ because otherwise $i$ can either benefit
or save money by switching to $z_{i}^{\prime }$ (property (\ref{12})). Since 
$z^{\prime }$ dominates $z$, some agent $j$ strictly prefers $z_{j}^{\prime
} $ to $z_{j}$, and therefore $z_{j}^{\prime }$ is outside his budget set,
i.e., $(p,z_{j}^{\prime })>(p,z_{j})$. Summing up these inequalities over
all agents we get the contradiction $(p,\omega )>(p,\omega )$. The argument
for the Weak Core property is similar. $\blacksquare \smallskip $

\textit{Remark 1:} \textit{A competitive allocation may fail the standard
Core from Equal Split property, where coalition }$S$\textit{\ blocks
allocation }$z$\textit{\ if it can use its endowment }$\frac{|S|}{n}e^{A}$%
\textit{\ to make everyone in }$S$\textit{\ weakly better off and at least
one agent strictly more. This is because \textquotedblleft equal
split\textquotedblright\ may give resources to agents who have no use for
them. Say three agents share one unit of item }$a$\textit{\ with }$%
u_{i}(z_{i})=z_{i}$\textit{\ for }$i=1,2$\textit{\ and }$u_{3}(z_{3})=-z_{3}$%
\textit{. The competitive allocation splits }$a$\textit{\ equally between
agents }$1$\textit{\ and }$2$\textit{, which coalition }$\{1,3\}$\textit{\
blocks by giving }$\frac{2}{3}$\textit{\ of }$a$\textit{\ to agent }$1$%
\textit{.}

\section{Main result}

We define formally the partition of division problems alluded to in the
Introduction. Given a problem $\mathcal{\mathcal{P}}$ we partition $N$ as
follows:%
\begin{equation*}
N_{+}=\{i\in N|\exists z\in \mathcal{F}:u_{i}(z_{i})>0\}\text{ ; }%
N_{-}=\{i\in N|\forall z\in \mathcal{F}:u_{i}(z_{i})\leq 0\}
\end{equation*}%
We call agents in $N_{+}$ \textit{attracted to} the manna, and those in $%
N_{-}$ \textit{repulsed by} it. All agents in $N_{-}$, and only those, are
globally satiated, and for them $z_{i}=0$ is a global maximum, not
necessarily unique.

The partition is determined by the relative position of the set $\mathcal{U}$
of feasible utility profiles and the cone $\Gamma =%
%TCIMACRO{\U{211d} }%
%BeginExpansion
\mathbb{R}
%EndExpansion
_{+}^{N_{+}}\times \{0\}^{N_{-}}$, where attracted agents benefit while
repulsed agents do not suffer. Let $\Gamma ^{\ast }=%
%TCIMACRO{\U{211d} }%
%BeginExpansion
\mathbb{R}
%EndExpansion
_{++}^{N_{+}}\times \{0\}^{N_{-}}$be the relative interior of $\Gamma $%
.\smallskip

\noindent \textbf{Lemma 2 }\textit{Each problem }$\mathcal{P}$ \textit{is of
(exactly) one of three types:}

\noindent \textit{positive if }$\mathcal{U}\cap \Gamma ^{\ast }\neq
\varnothing $; \textit{negative if }$\mathcal{U}\cap \Gamma =\varnothing $; 
\textit{null if }$\mathcal{U}\cap \Gamma =\{0\}$.\smallskip

Given a smooth function $f$ and a closed convex $C$ we say that $x\in C$%
\textit{\ is a critical point of }$f$\textit{\ in }$C$ if the upper contour
of $f$ at $x$ has a supporting hyperplane that supports $C$ as well:%
\begin{equation}
\forall y\in C:\partial f(x)\cdot y\leq \partial f(x)\cdot x\text{ and/or }%
\forall y\in C:\partial f(x)\cdot y\geq \partial f(x)\cdot x  \label{33}
\end{equation}%
This holds in particular if $x$ is a local maximum or local minimum of $f$
in $C$.\smallskip

In the next statement we write $\mathcal{U}^{eff}$ for the set of efficient
utility profiles, and $%
%TCIMACRO{\U{211d} }%
%BeginExpansion
\mathbb{R}
%EndExpansion
_{=}^{N}$ for the interior of $%
%TCIMACRO{\U{211d} }%
%BeginExpansion
\mathbb{R}
%EndExpansion
_{-}^{N}$.\smallskip

\noindent \textbf{Theorem }\textit{Competitive divisions exist in all
problems} $\mathcal{P}$.\textit{\ Moreover}

\noindent $i)$\textit{\ If }$\mathcal{P}$\textit{\ is positive their budget
is }$+1$;\textit{\ an allocation is competitive iff its utility profile
maximizes the product }${\Large \Pi }_{N_{+}}U_{i}$\textit{\ over }$\mathcal{%
U}\cap \Gamma ^{\ast }$\textit{; so }$CU(\mathcal{P)}$ \textit{contains a
single utility profile, positive in }$N_{+}$ \textit{and null in }$N_{-}$%
\textit{.}

\noindent $ii)$ \textit{If }$\mathcal{P}$ \textit{is negative their budget
is }$-1$\textit{; an allocation is competitive iff its utility profile is in 
}$\mathcal{U}^{eff}\cap R_{=}^{N}$\textit{\ and is a critical point of the
product }$\Pi _{N}|U_{i}|$\textit{\ in }$\mathcal{U}$\textit{; so all
utility profiles in }$CU(\mathcal{P)}$ \textit{are negative.}

\noindent $iii)$ \textit{If }$\mathcal{P}$\textit{\ is null their budget is }%
$0$\textit{;} \textit{an allocation is competitive iff its utility profile
is }$0$\textit{.\smallskip }

We see that the competitive utility profiles are entirely determined by the
set of feasible utility profiles: the competitive approach still has a
welfarist interpretation when we divide a mixed manna.

Moreover the Theorem implies that the task of dividing the manna is either
good news (at least weakly) for everyone, or strictly bad news for everyone.

The possible multiplicity of $CU(\mathcal{P)}$ for negative problems with
linear preferences is the subject of Subsection 6.1. Without backing up this
proposal by specific normative arguments, we submit that a natural selection
of $CU(\mathcal{P})$ obtains by maximizing the Nash product of individual 
\textit{dis}utilities on the negative efficiency frontier.\footnote{%
Note that minimizing the $\prod_{i\in N}|U_{i}|$ on $\mathcal{U}\cap \mathbb{%
R}_{-}^{N}$ picks a boundary point where this product is null, not a
competitive allocation.}\smallskip

\noindent \textbf{Lemma 3 }\textit{If }$\mathcal{P}$ \textit{is a negative
problem, the profile }$U^{\ast }$\textit{\ maximizing the Nash product }$%
\prod_{i\in N}|U_{i}|$\textit{\ over} $\mathcal{U}^{eff}\cap \mathbb{R}%
_{-}^{N}$ \textit{is a critical point of the product on }$\mathcal{U}$ 
\textit{and} $U_{i}^{\ast }<0$ \textit{for all }$i\in N$\textit{; hence} $%
U^{\ast }\in CU(\mathcal{P})$.\smallskip

This selection is almost always unique: we prove this in the linear
domain.\smallskip

\noindent \textbf{Lemma 4} \textit{Fix }$N$\textit{, }$A$\textit{\ and }$%
\omega $\textit{. For almost all negative problems} $\mathcal{P}%
=(N,A,u,\omega )$ \textit{with additive utilities (w.r.t. the Lebesgue
measure on the space }$R^{N\times A}$\textit{\ of utility matrices) the
utility profile }$U^{\ast }$\textit{\ defined in Lemma~3 is unique}%
.\smallskip

\textit{Remark 2 The Competitive Equilibrium with Fixed Income Shares (CEFI
for short)\ replaces in Definition 1 the common budget }$\beta $\textit{\ by
individual budgets }$\theta _{i}\beta $\textit{, where the positive weights }%
$\theta _{i}$\textit{\ are independent of preferences. It is well known that
in an all goods problem, this asymmetric generalization of the competitive
solution obtains by maximizing the weighted product }$\Pi _{N}U_{i}^{\theta
_{i}}$\textit{\ of utilities, so that it preserves the uniqueness,
computational and continuity properties of the symmetric solution. The same
is true of our Theorem that remains valid word for word for the CEFI
divisions upon raising }$U_{i}$\textit{\ to the power }$\theta _{i}$\textit{%
. In particular the partition of problems in positive, negative or null is
unchanged.}

\section{Additive utilities}

A utility function is now a vector $u_{i}\in 
%TCIMACRO{\U{211d} }%
%BeginExpansion
\mathbb{R}
%EndExpansion
^{A}$ and corresponding utilities are $U_{i}=u_{i}\cdot
z_{i}=\sum_{A}u_{ia}z_{ia}$. For agent $i$ item $a$ is a good (resp. a bad)
if $u_{ia}>0$ (resp. $u_{ia}<0$); if $u_{ia}=0$ she is satiated with any
amount of $a$. Given a problem $\mathcal{P}$ the following partition of
items is key to understanding the competitive divisions.

\begin{equation}
A_{+}=\{a|\exists i:u_{ia}>0\}\text{ ; }A_{-}=\{a|\forall i:u_{ia}<0\}\text{
; }A_{0}=\{a|\max_{i}u_{ia}=0\}  \label{a5}
\end{equation}%
We call an item in $A_{+}$ a \textit{collective good}, one in $A_{-}$ a 
\textit{collective bad}, and one in $A_{0}$ a \textit{neutral }item\textit{. 
}In an efficient allocation an item in $A_{+}$ is consumed only by agents
for whom it is a good, and a neutral item in $A_{0}$ is consumed only by
agents who are indifferent to it. We note that the above partition
determines the sign of competitive prices.

\textit{Fact: if }$(z,p,\beta )$\textit{\ is a competitive division, we have}%
\begin{equation}
p_{a}>0\text{ if }a\in A_{+}\text{ ; }p_{a}<0\text{ if }a\in A_{-}\text{ ; }%
p_{a}=0\text{ if }a\in A_{0}  \label{2}
\end{equation}%
The proof is simple. If the first statement fails an agent who likes $a$
would demand an infinite amount of it; if the second fails no one would
demand $b$. If the third fails with $p_{a}>0$ the only agents who demand $a$
have $u_{ia}=0$, so that eating some $a$ violates (\ref{12}); if it fails
with $p_{a}<0$ an agent such that $u_{ia}=0$ gets an arbitrarily cheap
demand by asking large amounts of $a$, so (\ref{12}) fails again.

\subsection{The multiplicity issue}

\noindent \textbf{Proposition 1 }\textit{If utilities are additive in
problem }$\mathcal{P}$, \textit{the number }$|CU(\mathcal{P})|$ \textit{of
distinct competitive utility profiles is finite. Set }$n=|N|$ and $m=|A|$%
\textit{, then}

\noindent $i)$ \textit{If }$n=2$ t\textit{he upper bound of }$|CU(\mathcal{P}%
)|$\textit{\ is }$2m-1$.

\noindent $ii)$ \textit{If }$m=2$ t\textit{he upper bound of }$|CU(\mathcal{P%
})|$\textit{\ is }$2n-1$.

\noindent $iii)$ \textit{For general }$n,m$,\textit{\ }$|CU(\mathcal{P})|$ 
\textit{can be as high as }$2^{\min \{n,m\}}-1$\textit{\ if }$n\neq m$%
\textit{, and }$2^{n-1}-1$\textit{\ if} $n=m$.\textit{\smallskip }

\noindent We offer no guess about the upper bound of $|CU(\mathcal{P})|$ for
general $n,m$.

Three examples follow to illustrate the Proposition. For statement $i)$ the
agents in $N=\{1,2\}$ share five bads $A=\{a,b,c,d,e,f\}$, one unit of each;
utilities are%
\begin{equation*}
\begin{array}{ccccccc}
& a & b & c & d & e & f \\ 
u_{1} & -1 & -1 & -2 & -4 & -8 & -17 \\ 
u_{2} & -17 & -8 & -4 & -2 & -1 & -1%
\end{array}%
\end{equation*}%
Here $|CE(\mathcal{P})|=|CU(\mathcal{P})|=11$. In five competitive
allocations no bad is split between the agents; agent $1$ eats all the bads
in a left interval of $A$, and agent $2$ all those in the complement right
interval of $A$. For instance $\{a,b\}$ for $1$ and $\{c,d,e,f\}$ for $2$ is
sustained by the price $p=-(\frac{1}{2},\frac{1}{2},\frac{1}{2},\frac{1}{4},%
\frac{1}{8},\frac{1}{8})$ and $\beta =-1$. In addition we have six
competitive allocations where exactly one bad is shared between $1$ and $2$,
while $1$ gets the bads to its left, if any, and $2$ those to its right, if
any. For instance if we split $f$ agent $1$ gets the five other bads and $%
\frac{1}{34}$ of $f$, while $2$ eats $\frac{33}{34}$ of $f$; the price is $%
p=-\frac{1}{33}(2,2,4,8,16,34)$. Notice that agent $1$ gets exactly his Fair
Share utility (from eating $\frac{1}{2}$ of every item).

For statement $ii)$ we take $N=\{1,2,3,4,5,6\}$, two bads $A=\{a,b\}$, one
unit of each, and the utilities%
\begin{equation*}
\begin{array}{ccccccc}
& u_{1} & u_{2} & u_{3} & u_{4} & u_{5} & u_{6} \\ 
a & -1 & -1 & -2 & -3 & -3 & -6 \\ 
b & -6 & -3 & -3 & -2 & -1 & -1%
\end{array}%
\end{equation*}%
Again $|CE(\mathcal{P})|=|CU(\mathcal{P})|=11$. The five allocations where
the left-most agents divide $a$ equally and eat no $b$, while the right-most
ones divide $b$ equally and eat no $a$, are competitive. For instance $1$
and $2$ share $a$ while $3,4,5,6$ share $b$ corresponds to $p=-(2,4)$ and $%
\beta =-1$. In the other six competitive divisions one agent eats some of
both bads, agents to his left eat only $a$ and agents to his right only $b$.
For instance the allocation%
\begin{equation*}
\begin{array}{ccccccc}
& z_{1} & z_{2} & z_{3} & z_{4} & z_{5} & z_{6} \\ 
a & 5/12 & 5/12 & 1/6 & 0 & 0 & 0 \\ 
b & 0 & 0 & 1/6 & 5/18 & 5/18 & 5/18%
\end{array}%
\end{equation*}%
is sustained by the price $p=-(\frac{12}{5},\frac{18}{5})$.

Finally for statement $iii)$ we set $N=\{1,2,3,4,5,6\}$, $A=\{a,b,c,d,e\}$,
one unit of each bad, and the utilities%
\begin{equation*}
\begin{array}{cccccc}
& a & b & c & d & e \\ 
u_{1} & -1 & -3 & -3 & -3 & -3 \\ 
u_{2} & -3 & -1 & -3 & -3 & -3 \\ 
u_{3} & -3 & -3 & -1 & -3 & -3 \\ 
u_{4} & -3 & -3 & -3 & -1 & -3 \\ 
u_{5} & -3 & -3 & -3 & -3 & -1 \\ 
u_{6} & -1 & -1 & -1 & -1 & -1%
\end{array}%
\end{equation*}%
We check that $|CE(\mathcal{P})|=|CU(\mathcal{P})|=31$. The symmetric
competitive division with uniform price $\frac{6}{5}$ for each bad gives to
each of the first five agents $\frac{5}{6}$ units of her preferred bad,
while agent $6$ eats $\frac{1}{6}$ of every bad, precisely his Fair Share.
Now for each strict subset of the first five agents, for instance $\{3,4,5\}$%
, there is a competitive allocation where each such agent eats
\textquotedblleft his\textquotedblright\ bad in full, while agent $1$ shares
the rest with the other agents:%
\begin{equation*}
\begin{array}{cccccc}
& a & b & c & d & e \\ 
z_{1} & 2/3 & 0 & 0 & 0 & 0 \\ 
z_{2} & 0 & 2/3 & 0 & 0 & 0 \\ 
z_{3} & 0 & 0 & 1 & 0 & 0 \\ 
z_{4} & 0 & 0 & 0 & 1 & 0 \\ 
z_{5} & 0 & 0 & 0 & 0 & 1 \\ 
z_{6} & 1/3 & 1/3 & 0 & 0 & 0%
\end{array}%
\end{equation*}%
Here prices are $p=-(\frac{3}{2},\frac{3}{2},1,1,1)$. This construction can
be adjusted for each non trivial partition of the first five agents. Note
that agent $6$'s utility goes from $-1$ (his Fair Share) to $-\frac{1}{2}$,
when he shares a single bad with a single other agent; utilities of other
agents vary also between $-1$ and $-\frac{1}{2}$.\smallskip

\textit{Remark 3. It is easy to show that for }$n=2$\textit{\ and/or} $m=2$, 
$|CU(\mathcal{P})|$ \textit{is odd in almost all problems (excluding only
those where the coefficients of }$u$\textit{\ satisfy certain simple
equations). We conjecture that a similar statement holds for any }$n,m$.

\subsection{Independence of Lost Bids}

We offer a compact axiomatic characterization of competitive fair division%
\textit{. }Because our axioms compare the selected allocations across
different problems, we define first division rules. Notation: when we
rescale each utility $u_{i}$ as $\lambda _{i}u_{i}$, the new utility matrix
is written $\lambda \ast u$.\smallskip

\noindent \textbf{Definition 2 }\textit{A division rule }$f$\textit{\
associates to every problem }$\mathcal{P}=(N,A,u,\omega )$\textit{\ a set\
of feasible allocations }$f(\mathcal{P})\subset \mathcal{F}(N,A,\omega )$%
\textit{\ such that for any rescaling }$\lambda $,\textit{\ }$\lambda _{i}>0$%
\textit{\ for all }$i$\textit{, we have: }$f(N,A,\lambda \ast u,\omega
)=f(N,A,u,\omega )$. \textit{Moreover }$f$\textit{\ meets
Pareto-Indifference (PI). For every }$\mathcal{P}$\textit{\ and }$%
z,z^{\prime }\in \mathcal{F}(N,A,\omega )$%
\begin{equation*}
\{z\in f(\mathcal{P})\text{ and }u_{i}\cdot z_{i}=u_{i}\cdot z_{i}^{\prime }%
\text{ for all }i\}\Longrightarrow z^{\prime }\in f(\mathcal{P})
\end{equation*}%
Note that PI implies that $f$ is entirely determined by its utility
correspondence $F(\mathcal{P})=\{u\cdot z|z\in f(\mathcal{P})\}$. The
invariance to rescaling property makes sure that division rules are ordinal
constructs, they only depend upon the underlying linear preferences.

The competitive division rule $\mathcal{P\rightarrow }CE(\mathcal{P)}$ meets
Definition 2. We give other examples after Proposition 2. Definition 2 is
not restricted to linear preferences, but our next axiom is.\smallskip

\noindent \textbf{Definition 3 }\textit{The division rule }$f$ \textit{is
Independent of Lost Bids (ILB) if for any two problems }$\mathcal{P},%
\mathcal{P}^{\prime }$ on $N,A,\omega $\textit{\ where }$u,u^{\prime }$ 
\textit{are additive, differ only in the entry }$ia$\textit{,\ and }$%
u_{ia}^{\prime }<u_{ia}$\textit{, we have}%
\begin{equation}
\forall z\in f(\mathcal{P}):z_{ia}=0\Longrightarrow z\in f(\mathcal{P}%
^{\prime })  \label{24}
\end{equation}%
Recall from Section 2 our interpretation of $u_{ia}$ as agent $i$'s bid for
item $a$. ILB says that the bid $u_{ia}$ only matters if it is winning, i.
e., agent $i$ eats some of item $a$. It can be shown that for a generic
utility matrix $u$ an efficient allocation $z$ has no more than $n+m-1$ non
zero coordinates (see Lemma 1 in \cite{BMSY}): then ILB reduces considerably
the number of parameters relevant to describe the outcome selected by the
rule.

That the competitive rule $\mathcal{P}\rightarrow CE(\mathcal{P)}$ meets ILB
is clear by Definition 1: as $a$ becomes less attractive to $i$ in the shift
from $\mathcal{P}$ to $\mathcal{P}^{\prime }$, $i$'s Walrasian demand can
only shrink, and it still contains $z_{i}$.

The characterization requires the uncontroversial fairness property known as 
\textbf{Equal Treatment of Equals} (ETE): for all $\mathcal{P}$%
\begin{equation*}
u_{i}=u_{j}\Longrightarrow U_{i}=U_{j}\text{ for all }U\in F(\mathcal{P})%
\text{ and all }i,j\in N
\end{equation*}%
We also impose the solidarity property uncovered in our Theorem. \textbf{%
Solidarity} (SOL): for all $\mathcal{P}$%
\begin{equation*}
U_{i}\cdot U_{j}\geq 0\text{ for all }U\in F(\mathcal{P})\text{ and all }%
i,j\in N
\end{equation*}%
Finally we call the rule $f$\ \textbf{Efficient} (EFF) if it selects only
efficient allocations in every problem\textit{\ }$\mathcal{P}$.\smallskip

\noindent \textbf{Proposition 2 }\textit{If a division rule meets Equal
Treatment of Equals, Solidarity, Efficiency and Independence of Lost Bids,
it contains the competitive rule.\smallskip }

\noindent If problem $\mathcal{P}$ involves only goods ($u_{ia}\geq 0$ for
all $i,a$) or only bads ($u_{ia}\leq 0$ for all $i,a$), Solidarity is
automatically true, so the characterization boils down to ETE, EFF and ILB.

We show after the proof (Subsection 7.7) that ILB is a strictly weaker
requirement than Maskin Monotonicity in the linear domain, thus connecting
Proposition 2 to earlier results mentioned in point 4 of Section 3.\smallskip

We discuss\textit{\ }the tightness of our characterization.

\noindent \textit{Drop ETE}. The CEFI division rule (Remark 1 Section 5)
fails ETE for general weights. It is straightforward to check that it meets
ILB either by suitably adapting Lemma 6 or directly in the general
Definition 1. Solidarity follows from our (adapted) Theorem.

\noindent \textit{Drop ILB. }Inspired by the Kalai-Smorodinsky bargaining
solution we construct now an efficient welfare rule $F$ meeting SOL and ETE
but failing ILB. Observe that if $\mathcal{P}$ is positive we have $%
U_{i}^{\max }=\max_{U\in \mathcal{U}}U_{i}>0$ for all $i\in N_{+}$, and if $%
\mathcal{P}$ is negative $U_{i}^{\min }=\min_{U\in \mathcal{U}}U_{i}<0$. In
a positive problem the rule picks the unique efficient utility profile $U$
such that $\frac{U_{i}}{U_{i}^{\max }}$ is constant for $i\in N_{+}$, and $%
U_{i}=0$ in $N_{-}$; in a negative problem it picks the efficient profile
such that $\frac{U_{i}}{U_{i}^{\min }}$ is constant for all $i$; and the
null utility at a null problem.

We do not know if the statement is tight with respect to SOL, but recall
that SOL is not needed for all goods or all bads problems. We conjecture
that the statement is tight with respect to EFF. We know at least that we
cannot drop both EFF and SOL, because a constrained version of the
competitive rule, where we impose $\sum_{A}z_{ia}=\frac{1}{n}\sum_{A}\omega
_{ia}$ as in \cite{HZ}, satisfies ETE and ILB.

\subsection{Single-valued Efficient and Envy-Free rules}

In this section and the next we uncover some negative features of the
competitive division rule in negative problems. It will be enough to state
them for \textquotedblleft all bads\textquotedblright\ problems. The first
result follows from a careful analysis of the set $\mathcal{A}$ of efficient
and envy-free allocations in problems with two bads $a,b,$ and any number of
agents.\smallskip

\noindent \textbf{Lemma 5 }\textit{If we divide at least two bads between at
least three agents, there are problems }$\mathcal{P}$ \textit{where the set }%
$\mathcal{A}$ \textit{of efficient and envy-free allocations, and the
corresponding set of disutility profiles, have }$\lfloor \frac{2n+1}{3}%
\rfloor $ \textit{connected components.\smallskip }

In a two-agent problem (even with mixed manna), No Envy coincides with Fair
Share Guarantee, so the set $\mathcal{A}$\ is clearly connected.

The proof of Lemma 5 makes clear that in a problem with exactly two bads the
maximal number of connected components of $\mathcal{A}$ is indeed $\lfloor 
\frac{2n+1}{3}\rfloor $. But we have no clue about the maximal number of
components in general all-bads\ problems. Nor do we know the answer for the
division of goods: if we divide exactly two goods,\ one can easily check that%
\textit{\ }$\mathcal{A}$\textit{\ }is connected. But beyond this simple case
we do not know if $\mathcal{A}$ remains connected in every \textquotedblleft
all goods\textquotedblright\ problem.\smallskip

We call the division rule $f$\ \textbf{Continuous} (CONT) if for each choice
of $N$, $A$, the corresponding welfare rule $(N,A,u,\omega )\rightarrow
F(N,A,u,\omega )$\ is a continuous function of\ $u\in 
%TCIMACRO{\U{211d} }%
%BeginExpansion
\mathbb{R}
%EndExpansion
^{N\times A}$. If the division rule does not depend upon the units of items
in $A$,\footnote{%
That is, for each $\lambda >0$ the set $F(\mathcal{P})$ is unchanged if we
replace $\omega _{a}$ by $\lambda \omega _{a}$ and $u_{ia}$ by $\frac{1}{%
\lambda }u_{ia}$. Clearly $CU$ meets this property.} CONT implies that $%
\mathcal{P}\rightarrow F(\mathcal{P})$ is also continuous in $\omega \in 
%TCIMACRO{\U{211d} }%
%BeginExpansion
\mathbb{R}
%EndExpansion
_{+}^{A}$.

We call the rule $f$\ \textbf{Envy-Free} (EVFR) if $f(\mathcal{P)}$ contains
at least one envy-free allocation for every problem $\mathcal{P}$.\smallskip

\noindent \textbf{Proposition 3 }\textit{If}\textbf{\ }\textit{we divide at
least two bads between at least four agents, no single-valued rule can be
Efficient, Envy-Free and Continuous.\smallskip }

This incompatibility result is tight. The equal division rule, $F_{i}(%
\mathcal{P})=\{\frac{1}{n}u_{i}\cdot \omega \}$ for all $\mathcal{P}$, is
EVFR and CONT. A single-valued selection of the competitive rule $CU$ meets
EFF and EVFR. The Egalitarian rule defined at the end of the previous
subsection meets EFF and CONT.

\subsection{Resource Monotonicity}

Adding more of an item that everyone likes to the manna, or removing some of
one that everyone dislikes, should not be bad news to anyone: the agents
own\ the items in common and welfare should be comonotonic to ownership.
When this property fails someone has an incentive to sabotage the discovery
of new goods, or add new bads to the manna.

We say that problem $\mathcal{P}^{\prime }$ improves problem $\mathcal{P}$
on item $a\in A$ if they only differ in the amount of item $a$ and either $%
\{\omega _{a}\leq \omega _{a}^{\prime }$ and $u_{ia}\geq 0$ for all $i\}$ or 
$\{\omega _{a}\geq \omega _{a}^{\prime }$ and $u_{ia}\leq 0$ for all $i\}$.

\textbf{Resource Monotonicity}\textit{\ }(RM): if $\mathcal{P}^{\prime }$
improves upon $\mathcal{P}$ on item $a\in A$, then $F(\mathcal{P})\leq F(%
\mathcal{P}^{\prime })$\smallskip

\noindent \textbf{Proposition 4}

\noindent $i)$ \textit{With two or more agents and two or more bads, no
efficient single-valued rule can be Resource Monotonic and Guarantee} 
\textit{Fair Share} ($u_{i}\cdot z_{i}\geq \frac{1}{n}u_{i}\cdot \omega $)%
\textit{.}

\noindent $ii)$ \textit{The competitive\ rule to divide goods is Resource
Monotonic (as well as single-valued, efficient and GFS).\smallskip }

The proof of statement $i)$ is by means of a simple two-person, two-bad
example. Fix a rule $F$ meeting EFF, RM and GFS. Consider the problem $%
\mathcal{P}$ with $%
\begin{array}{ccc}
& a & b \\ 
u_{1} & -1 & -4 \\ 
u_{2} & -4 & -1%
\end{array}%
$ and $\omega =(1,1)$, and set $U=F(\mathcal{P})$. As $-(1,1)$ is an
efficient utility profile, one of $U_{1},U_{2}$ is at least $-1$, say $%
U_{1}\geq -1$. Now let $\omega ^{\prime }=(\frac{1}{9},1)$ and pick $%
z^{\prime }\in f(\mathcal{P}^{\prime })$. By GFS and feasibility:%
\begin{equation*}
-z_{2b}^{\prime }\geq u_{2}\cdot z_{2}^{\prime }\geq \frac{1}{2}u_{2}\cdot
\omega ^{\prime }=-\frac{13}{18}
\end{equation*}%
\begin{equation*}
\Longrightarrow z_{1b}^{\prime }\geq \frac{5}{18}\Longrightarrow u_{1}\cdot
z_{1}^{\prime }=U_{1}^{\prime }\leq -\frac{10}{9}<U_{1}
\end{equation*}%
contradicting RM. Extending this argument to the general case $n\geq 3,m\geq
2$ is straightforward.

We omit for brevity the proof of statement $ii)$, available in \cite{BMSY}
as well as in \cite{SS1} for the more general cake-division model. It
generalizes easily to positive problems, when we add a unanimous good to an
already positive problem.

We stress that this positive result applies only to the linear domain, it
does not extend to general homothetic, convex and monotonic preferences. On
the latter domain, precisely the same combination of axioms as in
Proposition 4 cannot be together satisfied: see \cite{MT} and \cite{TK} .
This makes the goods versus bads contrast in the case of linear preferences
all the more intriguing.

\section{Appendix: Proofs}

\subsection{Lemma 2}

The three cases are clearly mutually exclusive; we check they are
exhaustive. It is enough to show that if $\mathcal{U}$ intersects $\Gamma
_{\neq 0}=\Gamma \diagdown \{0\}$ then it intersects $\Gamma ^{\ast }$ as
well. Let $z\in \mathcal{F}$ be an allocation with $u(z)\in \Gamma _{\neq 0}$
and $i_{+}$ be an agent with $u_{i_{+}}(z_{i_{+}})>0$. Define a new
allocation $z^{\prime }$ with $z_{i_{+}}^{\prime }=z_{i_{+}}+\varepsilon
\sum_{j\neq i_{+}}z_{j}$ and $z_{j}^{\prime }=(1-\varepsilon )z_{j}$ for $%
j\neq i_{+}$. By continuity we can select a small $\varepsilon >0$ such that 
$u(z^{\prime })\in \Gamma _{\neq 0}$. By construction $z_{i_{+}a}^{\prime
}>0 $ for all $a\in A$.

For any $j\in N_{+}\setminus \{i_{+}\}$ we can find $y_{j}\in \mathbb{R}^{A}$
such that $u_{j}(z_{j}^{\prime }+\delta y_{j})>0$ for small $\delta >0$.
Indeed if $u_{j}(z_{j}^{\prime })$ is positive we can take $y_{j}=0$. And if 
$u_{j}(z_{j}^{\prime })=0$, assuming that $y_{j}$ does not exist implies
that $z_{j}^{\prime }$ is a local maximum of $u_{j}$. By concavity of $u_{j}$
it is then a global maximum as well, which contradicts the definition of $%
N_{+}$.

Consider an allocation $z^{\prime \prime }$: $z_{i_{+}}^{\prime \prime
}=z_{i_{+}}^{\prime }-\delta \sum_{j\in N_{+}\setminus \{i_{+}\}}y_{j}$, $%
z_{j}^{\prime \prime }=z_{j}^{\prime }+\delta y_{j}$ for $j\in
N_{+}\setminus \{i_{+}\}$ and $z_{k}^{\prime \prime }=z_{k}^{\prime }$ for $%
k\in N_{-}$. For small $\delta >0$ this allocation is feasible and yields
utilities in $\Gamma ^{\ast }$.

\subsection{Main Theorem}

Throughout the proof it is convenient to consider competitive divisions $%
(z,p,\beta )$ with arbitrary budgets $\beta \in \mathbb{R}$ (not only $\beta
\in \{-1,0,1\}$ ); this clearly yields exactly the same set of competitive
allocations $CE(\mathcal{P)}$ and utility profiles $CU(\mathcal{P)}$.

\subsubsection{Positive problems: statement $i)$}

Let $\mathcal{N}({V})=\prod_{i\in N_{+}}{V}_{i}$ be the Nash product of
utilities of the attracted agents. We fix a positive problem $\mathcal{P}$
and proceed in two steps.\smallskip

\noindent \textbf{Step 1.}\textit{\ If }$U$\textit{\ maximizes} $\mathcal{N}(%
{V})$ \textit{over}\textbf{\ }${V}\in \mathcal{U}\cap \mathit{\Gamma }^{\ast
}$ \textit{and} ${z}\in \mathcal{F}$ \textit{is such that}\textbf{\ }$U=u(z)$%
\textbf{, }\textit{then }$z$\textit{\ is a competitive allocation with
budget }$\beta >0$\textit{.}

\noindent Let $\mathcal{C}_{+}$ be the convex cone of all ${y}\in \mathbb{R}%
_{+}^{N\times A}$ with $u(y)\in \Gamma $. For any $\lambda >0$ put 
\begin{equation*}
\mathcal{C}_{\lambda }=\left\{ {y}\in \mathcal{C}_{+}\mid \mathcal{N}%
(u(y))\geq \lambda ^{|N_{+}|}\right\} .
\end{equation*}%
Since $\mathcal{P}$ is positive the set $\mathcal{C}_{\lambda }$ is
non-empty for any $\lambda >0$. Continuity and concavity of utilities imply
that $\mathcal{C}_{\lambda }$ is closed and convex. Homogeneity of utilities
give $\mathcal{C}_{\lambda }=\lambda \mathcal{C}_{1}$.

Set $\lambda ^{\ast }=\left( \mathcal{N}(U)\right) ^{\frac{1}{|N_{+}|}}$.
The set $\mathcal{C}_{\lambda }$ does not intersect $\mathcal{F}$ for $%
\lambda >\lambda ^{\ast }$, and $\mathcal{C}_{\lambda ^{\ast }}$ touches $%
\mathcal{F}$ at $z$.

\noindent \textit{Step 1.1} \textit{There exists a hyperplane }$H$\textit{\
separating} $\mathcal{F}$ \textit{from} $\mathcal{C}_{\lambda ^{\ast }}$.

\noindent Consider a sequence $\lambda _{n}$ converging to $\lambda ^{\ast }$
from above. Since $\mathcal{C}_{\lambda _{n}}$ and $\mathcal{F}$ are convex
sets that do not intersect, they can be separated by a hyperplane $H_{n}$.
The family $\{H_{n}\}_{n\in \mathbb{N}}$ has a limit point $H$. The
hyperplane $H$ separates $\mathcal{F}$ from $\mathcal{C}_{\lambda ^{\ast }}$
by continuity of $u$. Thus there exist $q\in \mathbb{R}^{N\times A}$ and $%
Q\in \mathbb{R}$ such that $\sum_{i,a}q_{ia}{y}_{ia}\leq Q$ for $y\in 
\mathcal{F}$ and $\sum_{i,a}q_{ia}{y}_{ia}\geq Q$ on $\mathcal{C}_{\lambda
^{\ast }}$. The coefficients $q_{ia}$ will be used to define the vector of
prices $p$.

By the construction ${z}$ maximizes $\mathcal{N}(u({y}))$ over $\mathcal{B}%
^{N}(q,Q)=\{{y}\in \mathcal{C}_{+}\mid \sum_{i,a}q_{ia}{y}_{ia}\leq Q\}$.
Think of the latter as a \textquotedblleft budget set with agent-specific
prices\textquotedblright .

Define the vector of prices $p$ by $p_{a}=\max_{i\in N}q_{ia}$ and $\mathcal{%
B}^{\ast }(p,Q)=\{{y}\in \mathcal{C}_{+}\mid \sum_{i}p\cdot {y}_{i}\leq Q\}$%
. We show now that we do not need agent-specific pricing.

\noindent \textit{Step 1.2} \textit{The allocation }$z$\textit{\ maximizes} $%
\mathcal{N}(u({y}))$ \textit{over} $y\in \mathcal{B}^{\ast }(p,Q)$.

\noindent It is enough to show the double inclusion ${z}\in \mathcal{B}%
^{\ast }(p,Q)\subset \mathcal{B}^{N}(q,Q)$. The second one is obvious since $%
\sum_{i,a}{y}_{ia}p_{a}\leq Q$ implies $\sum_{i,a}{y}_{ia}q_{ia}\leq Q$. Let
us check the first inclusion. Taking into account that $z\in \mathcal{F}$
and $\sum_{i,a}q_{ia}y_{ia}\leq Q$ for $y\in \mathcal{F}$, we get 
\begin{equation*}
\sum_{i}p\cdot {z}_{i}=\sum_{a}p_{a}\sum_{i}{z}_{ia}=\sum_{a}p_{a}=\sum_{a}%
\max_{i}q_{ia}=\max_{y\in \mathcal{F}}\sum_{i,a}q_{ia}y_{ia}\leq Q.
\end{equation*}

\noindent \textit{Step 1.3} $({z},p,\beta )$ \textit{is a competitive
division for some }$\beta >0$.

\noindent Consider an agent $i$ from $N_{+}$. Check that the bundle ${z}_{i}$
belongs to his competitive demand $d_{i}(p,\beta _{i})$, where $\beta
_{i}=p\cdot {z}_{i}$. Indeed if there exists ${z_{i}^{\prime }}\in \mathbb{R}%
_{+}^{A}$ such that $p\cdot {z}_{i}^{\prime }\leq \beta _{i}$ and $u_{i}({%
z_{i}^{\prime }})>u_{i}({z}_{i})$, then switching the consumption of agent $%
i $ from ${z}_{i}$ to ${z_{i}^{\prime }}$ gives an allocation in $\mathcal{B}%
^{\ast }(p,Q)$ and increases the Nash product, contradicting Step 1.2. Note
that $\beta _{i}>0$ for $i\in N_{+}$ because otherwise we can take ${%
z_{i}^{\prime }}=2{z}_{i}$. Check now that $z_{i}$ is parsimonious: it
minimizes $p\cdot {y}_{i}$ over $d_{i}(p,\beta _{i})$. If not, pick $%
y_{i}\in d_{i}(p,\beta _{i})$ with $p\cdot y_{i}<p\cdot z_{i}$, then for $%
\delta $ small enough and positive, the bundle $z_{i}^{\prime }=(1+\delta
)y_{i}$ meets $p\cdot {z}_{i}^{\prime }\leq \beta _{i}$ and $u_{i}({%
z_{i}^{\prime }})>u_{i}({z}_{i})$.

We use now the classic equalization argument (\cite{E}) to check that $\beta
_{i}$ does not depend on $i\in N_{+}$. We refer to the fact that the
geometric mean is below the arithmetic one as \textquotedblleft the
inequality of means\textquotedblright .

Assume $\beta _{i}\neq \beta _{j}$ and consider a new allocation ${z^{\prime
}}$, where the budgets of $i$ and $j$ are equalized: ${z_{i}^{\prime }}=%
\frac{\beta _{i}+\beta _{j}}{2\beta _{i}}{z}_{i}$ and ${z_{j}^{\prime }}=%
\frac{\beta _{i}+\beta _{j}}{2\beta _{j}}z_{j}$. This allocation belongs to $%
\mathcal{B}^{\ast }(p,Q)$ and homogeneity of utilities implies 
\begin{equation*}
\mathcal{N}(u({z^{\prime }}))=\mathcal{N}(U)\left( \frac{\beta _{i}+\beta
_{j}}{2\beta _{i}}\right) \left( \frac{\beta _{i}+\beta _{j}}{2\beta _{j}}%
\right) .
\end{equation*}%
Now the (strict) inequality of means gives $\frac{\beta _{i}+\beta _{j}}{2}>%
\sqrt{\beta _{i}\beta _{j}}$, therefore $\mathcal{N}(u({z^{\prime }}))>%
\mathcal{N}(U)$ contradicting the optimality of ${z}$. Denote the common
value of $\beta _{i}$ by $\beta $.

Turning finally to the repulsed agents we check that for any $i\in N_{-}$
there is no $z_{i}^{\prime }$ such that $u_{i}(z_{i}^{\prime })=0$ and $%
\beta _{i}^{\prime }=p\cdot {z}_{i}^{\prime }<p\cdot {z}_{i}=\beta _{i}$,
i.e., $i$ can not decrease his spending. Assuming that $z_{i}^{\prime }$
exists we can construct an allocation $z^{\prime }\in \mathcal{B}^{\ast
}(p,Q)$, where agent $i$ switches to $z_{i}^{\prime }$, consumption of other
agents from $N_{-}$ remains the same, and ${z_{j}^{\prime }}={z}_{j}\frac{%
N_{+}\beta +\beta _{i}-\beta _{i}^{\prime }}{N_{+}\beta }$ for $j\in N_{+}$.
In other words, money saved by $i$ are redistributed among positive agents.
By homogeneity $\mathcal{N}(u(z^{\prime }))>\mathcal{N}(U)$, contradiction.
A corollary is that $\beta _{i}$ must be zero: take $z_{i}^{\prime }=0$ if $%
\beta _{i}>0$, and $z_{i}^{\prime }=2z_{i}$ if $\beta _{i}$ is negative. At $%
{z}_{i}$ agent $i$ reaches his maximal welfare of zero. Therefore, if $i$
can afford $z_{i}$, then $z_{i}$ is in the demand set. Since the price $%
\beta _{i}$ of $z_{i}$ is zero, we conclude $z_{i}\in d_{i}(p,\beta )$. The
proof of Step 1.3 and of Step 1 is complete.\medskip

\noindent \textbf{Step 2} \textit{If }$(z,p,\beta )$\textit{\ is a
competitive division, then }$\beta >0$\textit{, and }$U=u(z)$\textit{\
belongs to} $\mathcal{U}\cap \mathit{\Gamma }^{\ast }$ \textit{and maximizes}
$\mathcal{N}$ \textit{over this set.}

\noindent Check first $\beta >0$. If $\beta \leq 0$ the budget set $%
B(p,\beta )$ contains ${z}_{i}$ and $2{z}_{i}$ for all $i$, therefore $%
u_{i}(2z_{i})\leq u_{i}(z_{i})$ implies $U_{i}\leq 0$. Then $U$ is
Pareto-dominated by any ${U^{\prime }}\in \mathcal{U}\cap \mathit{\Gamma }%
^{\ast }$, contradicting the efficiency of $z$ (Lemma~1).

Now $\beta >0$ implies $U$ belongs to $\mathit{\Gamma }^{\ast }$: every $%
i\in N_{+}$ has a $y_{i}$ with $u_{i}(y_{i})>0$ and can afford $\delta y_{i}$
for small enough $\delta >0$; every $i\in N_{-}$ can afford $y_{i}=0$, hence 
$u_{i}(z_{i})=0$ and $p\cdot {z}_{i}\leq 0$ (by (\ref{12})).

Consider $U^{\prime }=u(z^{\prime })$ that maximizes $\mathcal{N}$ over $%
\mathcal{U}\cap \mathit{\Gamma }^{\ast }$. For any $i\in N_{+}$ his spending 
$\beta _{i}^{\prime }=p\cdot {z}_{i}^{\prime }$ must be positive. Otherwise $%
\delta z_{i}^{\prime }\in B(p,\beta )$ for any $\delta >0$ and agent $i$ can
reach unlimited welfare. Similarly $\beta _{i}^{\prime }<0$ for $i\in N_{-}$
implies $\delta z_{i}^{\prime }\in d_{i}(p,\beta )$ for any $\delta >0$, so
the spending in $d_{i}(p,\beta )$ is arbitrarily low, in contradiction of
parsimony (\ref{12}).

For attracted agents $\frac{\beta }{\beta _{i}^{\prime }}z_{i}^{\prime }\in
B(p,\beta )$ gives $\frac{\beta }{\beta _{i}^{\prime }}U_{i}^{\prime
}=u_{i}\left( \frac{\beta }{\beta _{i}^{\prime }}z_{i}^{\prime }\right) \leq
U_{i}$. Therefore if $U$ is not a maximizer of $\mathcal{N}$, we have 
\begin{equation*}
\mathcal{N}(U)<\mathcal{N}(U^{\prime })\leq \mathcal{N}(U)\prod_{i\in N_{+}}%
\frac{\beta _{i}^{\prime }}{\beta }\Longrightarrow 1<\left( \prod_{i\in
N_{+}}\frac{\beta _{i}^{\prime }}{\beta }\right) ^{\frac{1}{|N_{+}|}}\leq 
\frac{\sum_{i\in N_{+}}\beta _{i}^{\prime }}{|N_{+}|\beta }
\end{equation*}%
where we use again the inequality of means. Now we get a contradiction from 
\begin{equation*}
\sum_{i\in N_{+}}\beta _{i}^{\prime }\leq \sum_{i\in N}\beta _{i}^{\prime
}=p\cdot {\omega }=\sum_{i\in N}p\cdot {z}_{i}\leq \sum_{i\in N_{+}}\beta
+\sum_{i\in N_{-}}0=|N_{+}|\beta
\end{equation*}

\subsubsection{Negative problems: statement $ii)$}

The proof is simpler because we do not need to distinguish agents from $%
N_{+} $ and $N_{-}$. We define the Nash product for negative problems by $%
\mathcal{N}({V})=\prod_{i\in N}|{V}_{i}|$ and focus now on its critical
points in $\mathcal{U}^{eff}$. We start by the variational characterization
of such points. If $V\in \mathbb{R}_{=}^{N}$ we have$\frac{\partial }{%
\partial {{V}_{i}}}\mathcal{N}({V})=\frac{1}{{V}_{i}}\mathcal{N}({V})$.
Therefore ${U}\in \mathcal{U}\cap \mathbb{R}_{=}^{N}$ is a critical point of 
$\mathcal{N}$ on $\mathcal{U}$ that lay on $\mathcal{U}^{eff}$ iff%
\begin{equation}
\sum_{i\in N}\frac{{U}_{i}^{\prime }}{|U_{i}|}\leq -|N|\text{ \ for all }{U}%
^{\prime }\in \mathcal{U}  \label{34}
\end{equation}%
The choice of the sign in this inequality is determined by Efficiency. Set $%
\varphi _{U}({U^{\prime }})=\sum_{i\in N}\frac{{U}_{i}^{\prime }}{|U_{i}|}$:
inequality (\ref{34}) says that $U^{\prime }=U$ maximizes $\varphi _{U}({%
U^{\prime }})$ on $\mathcal{U}$.

We fix a negative problem $\mathcal{P}$ and proceed in two steps.\smallskip

\noindent \textbf{Step 1.}\textit{\ If a utility profile} ${U}\in \mathcal{U}%
^{eff}\cap \mathbb{R}_{=}^{N}$ \textit{is a critical point of} $\mathcal{N}$ 
\textit{on} $\mathcal{U}$\textit{, then any} ${z}\in \mathcal{F}$ \textit{%
implementing }$U$\textit{\ is a competitive allocation with budget }$\beta
<0 $\textit{.}

\noindent By (\ref{34}) for any ${y}\in \mathcal{F}$ we have $\varphi _{{U}%
}(u({y}))\leq -|N|$. Define 
\begin{equation*}
\mathcal{C}_{\lambda }=\left\{ {y}\in \mathbb{R}_{+}^{N\times A}\mid \varphi
_{{U}}(u({y}))\geq \lambda \right\}
\end{equation*}%
For $\lambda \leq 0$ it is non-empty (it contains $0$), closed and convex.
For $\lambda >-|N|$ the set $\mathcal{C}_{\lambda }$ does not intersect $%
\mathcal{F}$ and for $\lambda =-|N|$ it touches $\mathcal{F}$ at ${z}$.
Consider a hyperplane $\sum_{i,a}q_{ia}{y}_{ia}=Q$ separating $\mathcal{F}$
from $\mathcal{C}_{-|N|}$ and fix the sign by assuming $\sum_{i,a}q_{ia}{y}%
_{ia}\leq Q$ on $\mathcal{F}$ (existence follows as in Step 1.1 for positive
problems). By the construction ${z}$ maximizes $\varphi _{{U}}(u({y}))$ on $%
\mathcal{B}^{N}(q,Q)=\{{y}\in \mathbb{R}_{+}^{N\times A}\mid \sum_{i,a}q_{ia}%
{y}_{ia}\leq Q\}$. Defining prices by $p_{a}=\max_{i\in N}q_{ia}$ and
mimicking the proof of Step 1.2 for positive problems we obtain that ${z}$
belongs to $\mathcal{B}^{\ast }(p,Q)=\{{y}\in \mathbb{R}_{+}^{N\times A}\mid
\sum_{i,a}p_{a}{y}_{ia}\leq Q\}$ and maximizes $\varphi _{{U}}(u({y}))$
there.

We check now that $z$ is a competitive allocation with negative budget. For
any agent $i\in N$ the bundle $z_{i}$ belongs to his demand $d_{i}(p,\beta
_{i})$ (as before $\beta _{i}=p\cdot {z}_{i}$). If not, $i$ can switch to
any $z_{i}^{\prime }\in B(p,\beta _{i})$ with $u_{i}(z_{i}^{\prime })>U_{i}$%
, thus improving the value of $\varphi _{{U}}$ and contradicting the
optimality of $z$. The maximal spending $\beta _{i}$ must be negative,
otherwise $i$ can afford $y_{i}=0$ and $u_{i}(z_{i})<u_{i}(y_{i})$. If there
is some $z_{i}^{\prime }\in d_{i}(p,\beta _{i})$ such that $p\cdot {z}%
_{i}^{\prime }<\beta _{i}$, the bundle $z_{i}^{\prime \prime }=\frac{\beta
_{i}}{p\cdot {z}_{i}^{\prime }}z_{i}^{\prime }$ is still in $B(p,\beta _{i})$
and $u_{i}(z_{i}^{\prime \prime })>U_{i}$: therefore $p\cdot z_{i}=\beta
_{i} $ and $z_{i}$ is parsimonious ((\ref{12})).

Finally, $\beta _{i}=\beta _{j}$ for all $i,j\in N$. If $\beta _{i}\neq
\beta _{j}$, we use an \textit{un}equalization argument dual to the one in
Step 1.3 for positive problems. Assume for instance $\beta _{i}>\beta
_{j}\Leftrightarrow |\beta _{i}|<|\beta _{j}|$ and define ${z^{\prime }}$
from $z$ by changing only ${z_{i}^{\prime }}$ to $\frac{1}{2}{z}_{i}$ and ${%
z_{j}^{\prime }}$ to $\frac{2\beta _{j}+\beta _{i}}{2\beta _{j}}{z}_{j}$.
Clearly $z^{\prime }\in \mathcal{B}^{\ast }(p,Q)$ and we compute%
\begin{equation*}
\varphi _{{U}}(u({z^{\prime }}))-\varphi _{{U}}(U)=-\frac{1}{2}-\frac{2\beta
_{j}+\beta _{i}}{2\beta _{j}}+2=\frac{1}{2}-\frac{\beta _{i}}{2\beta _{j}}>0
\end{equation*}%
But we showed that $z$ maximizes $\varphi _{{U}}(u({y}))$ in $\mathcal{B}%
^{\ast }(p,Q)$: contradiction.\medskip

\noindent \textbf{Step 2.} \textit{If }$(z,p,\beta )$\textit{\ is a
competitive division, then }$\beta <0$\textit{\ and the utility profile }$%
U=u(z)$\textit{\ is a critical point of the Nash product on }$U$\textit{\
that belongs to} $\mathcal{U}^{eff}\cap \mathbb{R}_{=}^{N}$.

\noindent Check first that $\beta <0$. If not each agent can afford $y_{i}=0$
so ${U}_{i}\geq 0$ for all $i$, which is impossible in a negative problem.
Assume next ${U}_{i}\geq 0$ for some $i$: we have $2z_{i}\in B(p,\beta )$, $%
u_{i}(2z_{i})\geq u_{i}(z_{i})$, and $p\cdot (2z_{i})<p\cdot z_{i}$, which
contradicts (\ref{3}) and/or (\ref{12}) in Definition 1. Therefore ${U}$
belongs to $\mathbb{R}_{=}^{N}$. Finally $p\cdot z_{i}<\beta $ would imply $%
u_{i}(z_{i})<u_{i}(\lambda z_{i})$ for $\lambda \in \lbrack 0,1[$, and $%
\lambda z_{i}\in B(p,\beta )$ for $\lambda $ close enough to $1$, a
contradiction. Summarizing we have shown $U\in \mathcal{U}^{eff}\cap \mathbb{%
R}_{=}^{N}$ and $p\cdot z_{i}=\beta <0$ for all $i.$

To prove that ${U}$ is a critical point it is enough to check that it
maximizes $\varphi _{U}(u(y))$ on $\mathcal{F}$. Fix $z^{\prime }\in 
\mathcal{F}$, set $U^{\prime }=u(z^{\prime })$ and $p\cdot {z}_{i}^{\prime
}=\beta _{i}^{\prime }$. To show $\varphi _{U}(U^{\prime })\leq \varphi
_{U}(U)$ we will prove%
\begin{equation}
U_{i}^{\prime }\leq \frac{\beta _{i}^{\prime }}{\beta }U_{i}\text{ for all }i
\label{35}
\end{equation}%
This holds if $\beta _{i}^{\prime }<0$ because $\frac{\beta }{\beta
_{i}^{\prime }}z_{i}^{\prime }\in B(p,\beta )$ so $\frac{\beta }{\beta
_{i}^{\prime }}U_{i}^{\prime }=u_{i}\left( \frac{\beta }{\beta _{i}^{\prime }%
}z_{i}^{\prime }\right) \leq U_{i}$. If $\beta _{i}^{\prime }\geq 0$ we set $%
z_{i}^{\prime \prime }=\alpha z_{i}^{\prime }+(1-\alpha )z_{i}$, where $%
\alpha >0$ is small enough that $p\cdot {z}_{i}^{\prime \prime }<0$. We just
showed (\ref{35}) holds for $u_{i}(z_{i}^{\prime \prime })$, therefore%
\begin{equation*}
u_{i}(z_{i}^{\prime \prime })\leq \frac{p\cdot z_{i}^{\prime \prime }}{\beta 
}U_{i}=\alpha \frac{\beta _{i}^{\prime }}{\beta }U_{i}+(1-\alpha )U_{i}.
\end{equation*}%
Concavity of $u_{i}$ gives $\alpha U_{i}^{\prime }+(1-\alpha )U_{i}\leq
u_{i}(z_{i}^{\prime \prime })$ and the proof of (\ref{35}) is complete. Now
we sum up these inequalities and reach the desired conclusion%
\begin{equation*}
\varphi _{U}(U^{\prime })=\sum_{i\in N}\frac{U_{i}^{\prime }}{|U_{i}|}\leq -%
\frac{\sum_{i\in N}\beta _{i}^{\prime }}{\beta }=-\frac{\sum_{i\in N}p\cdot {%
z}_{i}^{\prime }}{\beta }=-\frac{p\cdot {\omega }}{\beta }=-|N|=\varphi
_{U}(U)
\end{equation*}

\subsubsection{\textbf{Null problems: statement }$iii)$}

The proof resembles that for positive problems, as we must distinguish $%
N_{+} $ from $N_{-}$, but the Nash product no longer plays a role. Fix a
null problem $\mathcal{P}$.

\noindent \textbf{Step 1.} \textit{Any} ${z}\in \mathcal{F}$ \textit{such
that }$u(z)=0$\textit{\ is competitive with} $\beta =0$.

Suppose first all agents are repulsed, $N=N_{-}$. Then $u_{i}({y}_{i})\leq 0$
for all $i\in N$ and ${y}_{i}\in \mathbb{R}_{+}^{A}$ and $({z},0,0)$ is a
competitive division: everybody has zero money, all bundles are free and all
agents achieve the best possible welfare with the smallest possible
spending. We assume from now on $N_{+}\neq \varnothing $.

Define $\psi (y)=\min_{i\in N_{+}}u_{i}(y_{i})$ for $y\in \mathbb{R}%
_{+}^{N\times A}$ and the sets $\mathcal{C}_{\lambda }=\{y\in \mathcal{C}%
_{+}\mid \psi (y)\geq \lambda \}$, where $\mathcal{C}_{+}=\{{y}\in \mathbb{R}%
_{+}^{N\times A}|u(y)\in \Gamma \}$ (as in the positive proof). For $\lambda
\geq 0$ the set $\mathcal{C}_{\lambda }$ is non-empty, closed and convex. If 
$\lambda >0$, the sets $\mathcal{C}_{\lambda }$ and $\mathcal{F}$ do not
intersect. As in Step 1.1 of the positive proof we construct a hyperplane
separating $\mathcal{F}$ and $\mathcal{C}_{0}$, define the set $\mathcal{B}%
^{N}(q,Q)$, the vector of prices $p$, and the set $\mathcal{B}^{\ast }(p,Q)$%
. Similarly we check that the allocation $z$ maximizes $\psi (y)$ over $y\in 
\mathcal{B}^{\ast }(p,Q)$, and $\psi (z)=0$.

We set $\beta _{i}=p\cdot {z}_{i}$ and show that $(z,p,0)$ is a competitive
division in three substeps.

\noindent \textit{Step 1.1 for all }$i\in N$ \textit{and} $x_{i}\in 
%TCIMACRO{\U{211d} }%
%BeginExpansion
\mathbb{R}
%EndExpansion
_{+}^{A}$: $p\cdot x_{i}<\beta _{i}\Longrightarrow u_{i}(x_{i})<0$.

\noindent Suppose $p\cdot x_{i}<\beta _{i}$ and $u_{i}(x_{i})\geq 0$ for
some $i\in N_{+}$. For each $j$ in $N_{+}$ pick a bundle $y_{j}^{+}$ such
that $u_{j}(y_{j}^{+})>0$ and construct the allocation $z^{\prime }$ as
follows: $z_{i}^{\prime }=x_{i}+\delta y_{i}^{+}$; $z_{j}^{\prime }=$ $%
z_{j}+\delta y_{j}^{+}$ for any other $j\in N_{+}$; $z_{j}^{\prime }=z_{j}$
for $j\in N_{-}$. If $\delta >0$ is small enough $z^{\prime }\in \mathcal{B}%
^{\ast }(p,Q)$ and for any $j\in N_{+}$ we have $u_{j}(z_{j}^{\prime })>0$,
by concavity and homogeneity of $u_{j}$. For instance 
\begin{equation*}
\frac{1}{2}u_{i}(z_{i}^{\prime })=u_{i}\left( \frac{1}{2}x_{i}+\frac{1}{2}%
\delta y_{i}^{+}\right) \geq \frac{1}{2}u_{i}(x_{i})+\frac{1}{2}\delta
u_{i}(y_{i}^{+})>0
\end{equation*}%
Therefore $\psi (z^{\prime })>0$ contradicting the optimality of $z$.

The proof when $p\cdot x_{i}<\beta _{i}$ and $u_{i}(x_{i})\geq 0$ for some $%
i\in N_{-}$ is similar and left to the reader.

\noindent \textit{Step 1.2 }$\beta _{i}=0$\textit{\ for all }$i\in N$.

\noindent If $\beta _{i}>0$ then $x_{i}=0$ is such that $p\cdot x_{i}<\beta
_{i}$ and $u_{i}(x_{i})=0$, which we just ruled out. If $\beta _{i}<0$ then $%
p\cdot (2z_{i})<\beta _{i}$ yet $u_{i}(2z_{i})=0$, contradicting Step 1.1.

From Steps 1.1, 1.2 we see that for all $i$ if $y_{i}\in d_{i}(p,0)$ then $%
p\cdot y_{i}=0$: so if we show $z_{i}\in d_{i}(p,0)$ the parsimony property (%
\ref{12}) is automatically satisfied. Therefore our next substep completes
the proof of Step 1.

\noindent \textit{Step 1.3} $z_{i}\in d_{i}(p,0)$ \textit{for all} $i\in N$.

For $i\in N_{-}$ this is obvious since such agent reaches his maximal
welfare $u_{i}=0$. Pick now $i\in N_{+}$ and assume $z_{i}\not\in d_{i}(p,0)$%
. Then $d_{i}(p,0)$ contains some $y_{i}$ with $u_{i}(y_{i})>0$. Let $w$ be
a bundle with negative price. Such bundle exists since $p\cdot \omega
=\sum_{i\in N}\beta _{i}=0$ and $p\neq 0$. Hence the bundle $%
x_{i}=y_{i}+\delta w$ with small enough $\delta >0$ has negative price $%
p\cdot x_{i}<0$ and $u_{i}(x_{i})>0$. Contradiction.\medskip

\noindent \textbf{Step 2.} \textit{If }$(z,p,\beta )$\textit{\ is a
competitive division, then }$u(z)=0$\textit{\ and }$(z,p,\beta ^{\prime })$%
\textit{\ with }$\beta ^{\prime }=0$\textit{\ is also competitive.}

\noindent If $\beta <0$ we have $u_{i}({z}_{i})<0$ for all $i\in N$.
Otherwise $u_{i}(z_{i})\geq 0$ and $p\cdot {z}_{i}<0$ implies as before that 
${z}_{i}^{\prime }=2{z}_{i}$ improves $U_{i}$ (at least weakly) while
remaining in $B(p,\beta )$ and lowering $i$'s spending. But $U\in 
%TCIMACRO{\U{211d} }%
%BeginExpansion
\mathbb{R}
%EndExpansion
_{=}^{N}$ is not efficient in a null problem.

Thus $\beta \geq 0$, hence $u_{i}({z}_{i})\geq 0$ for all $i\in N$ because
the bundle $0$ is in the budget set. The problem is null therefore $u({z})=0$%
, implying $0\in d_{i}(p,\beta )$ and by parsimony (\ref{12}) $p\cdot {z}%
_{i}\leq 0$, for all $i$. Hence $z_{i}\in d_{i}(p,0)$ therefore $(z,p,0)$ is
clearly a competitive division.

\subsection{Lemma 3}

We have $u_{i}(\omega )<0$ for every $i\in N$, else the allocation $z$ with $%
z_{i}=\omega $ and $z_{j}=0$ for $j\neq i$ yields utilities in $\Gamma $.

Consider the set of utility profiles dominated by $\mathcal{U}\cap \mathbb{R}%
_{-}^{N}$: $\mathcal{U}_{\leq }=\{U\in \mathbb{R}_{-}^{N}|\exists U^{\prime
}\in \mathcal{U}\cap \mathbb{R}_{-}^{N}:U\leq U^{\prime }\}$. This set is
closed and convex and contains all points in $\mathbb{R}_{-}^{N}$ that are
sufficiently far from the origin. Indeed, any $U\in \mathbb{R}_{-}^{N}$ such
that $U_{N}<\min_{i}u_{i}(\omega )$, where $U_{N}=\sum_{i}U_{i}$, is
dominated by the utility profile $z:z_{i}=\frac{U_{i}}{U_{N}}\omega ,\ i\in
N $.

Fix $\lambda \geq 0$ and consider the upper contour of the Nash product at $%
\lambda $: $C_{\lambda }=\{U\in \mathbb{R}_{-}^{N}\mid {\Large \Pi }%
_{N}|U_{i}|\geq \lambda \}$. For sufficiently large $\lambda $ the closed
convex set $C_{\lambda }$ is contained in $\mathcal{U}_{\leq }$. Let $%
\lambda ^{\ast }$ be the minimal $\lambda $ with this property. Negativity
of $\mathcal{P}$ implies that $\mathcal{U}_{\leq }$ is bounded away from $0$
so that $\lambda ^{\ast }$ is strictly positive. By definition of $\lambda
^{\ast }$ the set $C_{\lambda ^{\ast }}$ touches the boundary of $\mathcal{U}%
_{\leq }$ at some $U^{\ast }$ with strictly negative coordinates. Let $H$ be
a hyperplane supporting $\mathcal{U}_{\leq }$ at $U^{\ast }$. By the
construction, this hyperplane also supports $C_{\lambda ^{\ast }}$,
therefore $U^{\ast }$ is a critical point of the Nash product on $\mathcal{U}%
_{\leq }$: that is, $U^{\ast }$ maximizes $\sum_{i\in N}\frac{U_{i}}{%
|U_{i}^{\ast }|}$ over all $U\in \mathcal{U}_{\leq }$. So $U^{\ast }$
belongs to the Pareto frontier of $\mathcal{U}_{\leq }$, which is contained
in the Pareto frontier of $\mathcal{U}$. Thus $U^{\ast }$ is a critical
point of the Nash product on $\mathcal{U}$ and belongs to $\mathcal{U}%
^{eff}\cap \mathbb{R}_{=}^{N}$. By the construction any $U$ in the interior
of $C_{\lambda ^{\ast }}$ is dominated by some $U^{\prime }\in \mathcal{U}%
\cap \mathbb{R}_{-}^{N}$: so $U^{\ast }$ maximizes the Nash product on $%
\mathcal{U}^{eff}\cap \mathbb{R}_{-}^{N}$.

\subsection{Lemma 4}

In the previous proof note that the supporting hyperplane $H$ to $\mathcal{U}
$ at $U^{\ast }$ is unique because it is also a supporting hyperplane to $%
C_{\lambda ^{\ast }}$ that is unique. Hence, if $\mathcal{U}$ is a polytope
(e.g., for additive utilities), $U^{\ast }$ belongs to a face of maximal
dimension.

When utilities are additive, both sets $\mathcal{U}$ and $\mathcal{U}_{\leq
} $ are polytopes. Let $D\subset \mathbb{R}^{N\times A}$ be the set of all $%
u $ such that the problem $(N,A,u)$ is negative and $U^{\ast }$ is not
unique. By the above remark if $u\in D$ then for some $\lambda >0$ the set $%
\mathcal{U}_{\leq }$ has at least two faces $F$ and $F^{\prime }$ of maximal
dimension that are tangent to the surface $S_{\lambda }$: $\prod_{i\in
N}|U_{i}|=\lambda $, $U\in \mathbb{R}_{=}^{N}$. The condition that $F$ is
tangent to $S_{\lambda }$ fixes $\lambda $. The set of all hyperplanes
tangent to a fixed surface $S_{\lambda }$ has dimension $|N|-1$ (for every
point on $S$ there is one tangent hyperplane) though the set of all
hyperplanes in $\mathbb{R}^{N}$ is $|N|$-dimensional. Hence tangency of $%
F^{\prime }$ and $S_{\lambda }$ cuts one dimension. So $D$ is contained in a
finite union of algebraic surfaces and, therefore, has Lebesgue-measure zero.

\subsection{KKT conditions for additive utilities}

The first order characterization of the competitive allocations is very
useful in the proof of Propositions 1,2 and 3. Recall the partition of $A$ (%
\ref{a5}) and the correponding signs of the competitive prices (\ref{2}).

\noindent \textbf{Lemma 6}

\noindent $i)$ \textit{If }$\mathcal{P}$ \textit{is positive }$(z,p,+1)$%
\textit{\ is a competitive division iff }$p$ \textit{meets (\ref{2}) and:}

\textit{for all }$i\in N_{-}$\textit{: }$U_{i}=0$\textit{\ and }$p\cdot
z_{i}=0$\textit{;}

\textit{for all }$i\in N_{+}$\textit{: }$U_{i}>0$\textit{\ and }$p\cdot
z_{i}=1=\frac{1}{|N_{+}|}p\cdot \omega $\textit{; moreover}%
\begin{equation}
\text{for all }a\in A_{+}\cup A_{-}\text{\ }\{z_{ia}>0\}\Longrightarrow 
\frac{u_{ia}}{U_{i}}=p_{a}=\max_{j\in N_{+}}\frac{u_{ja}}{U_{j}}  \label{a1}
\end{equation}%
\noindent $ii)$ \textit{If }$\mathcal{P}$ \textit{is negative }$(z,p,-1)$%
\textit{\ is a competitive division iff for all }$i\in N$\textit{: }$U_{i}<0$%
\textit{\ and }$p\cdot z_{i}=-1=\frac{1}{|N|}p\cdot \omega $\textit{;
moreover}%
\begin{equation}
\text{for all }a\in A_{+}\cup A_{-}\text{\ }\{z_{ia}>0\}\Longrightarrow 
\frac{u_{ia}}{|U_{i}|}=p_{a}=\max_{j\in N}\frac{u_{ja}}{|U_{j}|}  \label{a2}
\end{equation}%
\noindent $iii)$ \textit{If} $\mathcal{P}$ \textit{is null }$z$ \textit{is a
competitive allocation iff }$U_{i}=0$ \textit{for all }$i\in N$\textit{.
Then }$(p,0)$ \textit{is a corresponding price and budget iff\ }$p\cdot
z_{i}=0=p\cdot \omega $\textit{\ and there exists }$\lambda \in 
%TCIMACRO{\U{211d} }%
%BeginExpansion
\mathbb{R}
%EndExpansion
_{++}^{N_{+}}$\textit{\ such that}%
\begin{equation}
\text{for all }a\in A_{+}\cup A_{-}\text{\ }\{z_{ia}>0\}\Longrightarrow
\lambda _{i}u_{ia}=p_{a}=\max_{j\in N_{+}}\lambda _{j}u_{ja}  \label{a3}
\end{equation}

\noindent A consequence of Lemma 6 is that for all $a\in A_{+}$ and $b\in
A_{-}$ and all $i\in N$ we have%
\begin{equation*}
\frac{u_{ia}}{p_{a}}\leq |U_{i}|\leq \frac{u_{ib}}{p_{b}}
\end{equation*}%
with equality on the left if $z_{ia}>0$ and on the right if $z_{ib}>0$. Also
an agent $i\in N_{-}$ eats only in $A_{0}$ if $\mathcal{P}$ is positive or
null, and only in $A_{-}\cup A_{0}$ if $\mathcal{P}$ is negative.

\textbf{Proof }\textit{Statement }$i)$. Assume $\mathcal{P}$ is positive.
Assume $(z,p,+1)$ is competitive: by our Theorem it maximizes the product of
utilities over $\mathcal{U\cap }\Gamma ^{\ast }$. Therefore $U_{j}=0$ in $%
N_{-}$ and $p\cdot z_{j}=0$ because such agent eats only in $A_{0}$ which is
free. Agent $i\in N_{+}$ clearly spends all his budget so $p\cdot z_{i}=0$.
If $z_{ia}>0$ and we transfer a vanishingly small amount of item $a$ to
agent $j\in N_{+}$, inequality $\frac{u_{ia}}{U_{i}}\geq \frac{u_{ja}}{U_{j}}
$ guarantees that the product of utilities does not increase; this implies $%
\frac{u_{ia}}{U_{i}}=\max_{j\in N_{+}}\frac{u_{ja}}{U_{j}}$. Now for any $%
a,b\in A_{+}\cup A_{-}$ such that $z_{ia}>0$ and $z_{ib}>0$ (\ref{3})
implies $\frac{u_{ia}}{p_{a}}=\frac{u_{ib}}{p_{b}}$. Multiplying numerator
and denominator by $z_{ia}$ and summing up over the support of $z_{i}$ in $%
A_{+}\cup A_{-}$ we get%
\begin{equation}
\frac{u_{ia}}{p_{a}}=\frac{\sum_{A_{+}\cup A_{-}}u_{ib}z_{ib}}{%
\sum_{A_{+}\cup A_{-}}p_{b}z_{ib}}=\frac{U_{i}}{1}  \label{a4}
\end{equation}%
where the second equality holds because extending the sum to $A_{0}$ changes
nothing because $p_{b}=0$ and $z_{ib}>0$ can only happen if $u_{ib}=0$. This
proves (\ref{a1}).

Conversely pick $(z,p,+1)$ meeting (\ref{a1}) and the two properties just
before. Items in $A_{+}\cup A_{-}$ are eaten exclusively by agents in $N_{+}$%
. Check that an item $a\in A_{0}$ can only be eaten by $i$ if $u_{ia}=0$:
for $i\in N_{-}$ this follows from $U_{i}=0$, and for $i\in N_{+}$ it
follows from writing (\ref{a1}) as $\frac{u_{ia}}{p_{a}}=U_{i}$ for $a\in
A_{+}\cup A_{-}$ such that $z_{ia}>0$ and summing up as in (\ref{a4}) to get 
$U_{i}=\sum_{A_{+}\cup A_{-}}u_{ib}z_{ib}$.

Therefore the maximization of $\Pi _{N_{+}}U_{i}$ in $\mathcal{U\cap }\Gamma
^{\ast }$ is equivalent to that of $\Pi _{N_{+}}u_{i}\cdot z_{i}$ when those
agents share the items in $A_{+}\cup A_{-}$, up to an arbitrary distribution
of $A_{0}$ to agents who do not mind. The KKT optimality conditions of this
latter problem obtain from (\ref{a1}) by ignoring the prices $p_{a}$. We
conclude by our Theorem that $(z,p,+1)$ is a competitive division.

\noindent \textit{Statement }$ii)$ The proof is similar and simpler, because
we do not have to distinguish between agents in $N_{+}$ and $N_{-}$. It is
omitted for brevity.

\noindent \textit{Statement }$iii)$ The first sentence, and the fact that
the competitive budget is $0$ with $p\cdot z_{i}=0$ for all $i$ come from
our Theorem. We check now that if $\mathcal{P}$ is null, and $z\in \mathcal{F%
}$ implements the zero utility profile, there exists $\lambda ,p$ meeting (%
\ref{a3}). Recall that $\mathcal{U}$ intersects $\Gamma =%
%TCIMACRO{\U{211d} }%
%BeginExpansion
\mathbb{R}
%EndExpansion
_{+}^{N_{+}}\times \{0\}^{N_{-}}$ only at $0$ (Lemma 2), and in fact $%
\mathcal{U\cap }%
%TCIMACRO{\U{211d} }%
%BeginExpansion
\mathbb{R}
%EndExpansion
_{+}^{N}=\{0\}$ as well because no agent in $N_{-}$ can get a positive
utility. As $\mathcal{U}$ is a polytope, we can separate it from $%
%TCIMACRO{\U{211d} }%
%BeginExpansion
\mathbb{R}
%EndExpansion
_{+}^{N}$ by a strictly positive vector $\lambda $. The separation property
is $\sum_{N}\lambda _{i}(u_{i}\cdot y_{i})\leq 0=\sum_{N}\lambda
_{i}(u_{i}\cdot z_{i})$ for any $y\in \mathcal{F}$, which implies $\lambda
_{i}u_{ia}=\max_{j\in N}\lambda _{j}u_{ja}$ whenever $z_{ia}>0$. So if we
define $p_{a}=\max_{j\in N}\lambda _{j}u_{ja}$ for all $a\in A_{+}\cup A_{-}$%
, property (\ref{a3}) follows at once. Clearly $p_{a}$ is null if $a\in
A_{0} $ and the price of other items is non zero: then (\ref{a3}) implies
that $z_{i}$ is agent $i$'s competitive demand at $p$ (property (\ref{3})).

We omit the simple proof of the converse statement: if $(z,p,0)$ is a
competitive division, there exists $\lambda $ meeting (\ref{a3}).

\subsection{Proposition 1}

\noindent \textit{Step 1. Check that} $CU(\mathcal{P)}$ \textit{is finite}.
If $\mathcal{P}$ is positive or null, this follows from the Theorem, even
without assuming linear preferences. If $\mathcal{P}$ is negative the set $B=%
\mathcal{U(P)\cap 
%TCIMACRO{\U{211d} }%
%BeginExpansion
\mathbb{R}
%EndExpansion
}_{-}^{N}$ is a polytope and the Theorem says that at each profile $U\in CU(%
\mathcal{P)}$ the function $\Pi _{N}|U_{i}|$ is critical in $B$. Clearly
this function has at most one critical point in the interior of each face of 
$B$, and there is a finite number of such interiors.\smallskip

\noindent \textit{Step 2. Statement }$iii)$. We generalize the numerical
example with 6 agents and 5 bads just after the Proposition to $n$ agents
and $m$ bads with $n>m$. We set $A=\{a_{1},\cdots ,a_{m}\}$ and use the
notation $e^{S}$ for the vector in $%
%TCIMACRO{\U{211d} }%
%BeginExpansion
\mathbb{R}
%EndExpansion
^{A}$ with $e_{i}^{S}=1$ if $i\in S$ and zero otherwise. For agent $i,1\leq
i\leq m,$ set as before $u_{ia_{i}}=-1,u_{ia_{j}}=-3$ for $j\neq i$, and for
agents $m+1$ to $n$ pick $u_{ia}=-1$ for all $a$. Then for any $q,1\leq
q\leq m$, the allocation%
\begin{equation*}
z_{i}=\frac{m}{n}e^{a_{i}}\text{ for }1\leq i\leq q\text{ ; }z_{j}=e^{a_{j}}%
\text{ for }q+1\leq j\leq m\text{ ; }z_{j}=\frac{1}{n}e^{\{a_{1},\cdots
,a_{q}\}}\text{ for }m+1\leq j\leq n
\end{equation*}%
is competitive for the prices $p_{a_{i}}=-\frac{q+1}{q}$ for $1\leq i\leq q$
and $p_{a_{j}}=-1$ for $q+1\leq j\leq m$.

Similarly in the case $m>n$ we set $u_{ia_{k}}=-1$ for $k=i$ or $n+1\leq
k\leq m$, and $u_{ia_{k}}=-3$ for $k\leq n,k\neq i$. Then for any subset of
agents $N^{\ast }\subseteq N$ the allocation where those agents share
equally the bads $a_{n+1},\cdots ,a_{m}$, while bad $a_{i},1\leq i\leq n$
goes to agent $i$, is competitive with prices $p_{a_{n+1}}=p_{a_{i}}=-\frac{%
n^{\ast }}{n^{\ast }+1}$ for $i\in N^{\ast }$, $p_{a_{j}}=-1$ for $j\in
N\diagdown N^{\ast }$.

This construction can be repeated for any subset of $m$ bads, thus
generating $2^{m}-1$ different competitive divisions. We omit for brevity
the similar argument for the case $m>n$.\smallskip

For the longer proof of the statements $i)$ and $ii)$ Lemma 6 is
critical.\smallskip

\noindent \textit{Step 3. Statement} $i)\smallskip $

We fix a negative problem $\mathcal{P}=(N=\{1,2\},A,u,\omega )$. If $A_{0}$
is non empty, it is easy to check that $CU(\mathcal{P})$ does not change if
we simply drop those items; so we assume from now on that $A_{+}$ and $A_{-}$
partition $A$, and $A_{-}$ is non empty. If $u_{1a}$ and $u_{2a}$ are of
strictly opposite signs for some item $a$, for instance $u_{1a}>0>u_{2a}$,
then by Efficiency item $a$ goes entirely to agent $1$ and if we replace $%
u_{2a}$ by $u_{2a}^{\prime }=0$ then again $CU(\mathcal{P})$ is unchanged,
so we can assume that all items in $A_{+}$ have $u_{ia}\geq 0$ for $i=1,2$
with at least one strictly (for an item with $u_{1a}=u_{2a}=0$ can be
discarded as well). For all items in $A_{-}$ we have as usual $u_{ia}<0$ for 
$i=1,2$. If $a\in A_{+}$ (resp. $A_{-}$) we say for clarity that $a$ is a
good (resp. a bad). We keep in mind that prices are positive for goods and
negative for bads.

We label the items $k\in \{1,\cdots ,m\}$ so that the ratios $\frac{u_{1k}}{%
u_{2k}}$ increase weakly in $k$. with the convention $\frac{1}{0}=\infty $.
We will prove the statement first when the sequence $\frac{u_{1k}}{u_{2k}}$
increases strictly in $k$.

\noindent \textit{Step 3.1} We fix a competitive division $(z,p,-1)$ and
show three properties of $z$:

\noindent $a)$ if $k,k^{\prime }$ are bads and $z_{1k},z_{2k^{\prime }}>0$
then $k<k^{\prime }$

\noindent $b)$ if $\ell ,\ell ^{\prime }$ are goods and $z_{1\ell },z_{2\ell
^{\prime }}>0$ then $\ell ^{\prime }<\ell $

\noindent $c)$ if $k$ is a bad, $\ell $ is a good, and $z_{1k},z_{1\ell }>0$
then $k<\ell $; if $z_{2k},z_{2\ell }>0$ then $\ell <k$

Condition $a)$, $b)$ follow directly from Efficiency. Pick $k,k^{\prime }$
as in the premises of $a)$ but such that $k^{\prime }<k$: then transferring $%
\varepsilon $ units of $k$ from $1$ to $2$, against $\delta $ units of $%
k^{\prime }$ from $2$ to $1$ (which is feasible for $\varepsilon ,\delta $
small enough) is beneficial to both if $\frac{u_{1k^{\prime }}}{u_{1k}}<%
\frac{\varepsilon }{\delta }<\frac{u_{2k^{\prime }}}{u_{2k}}$ which is
feasible as $k^{\prime }<k$. The argument for $b)$ is similar.

For condition $c)$ we must use the competitiveness assumption in particular
property (\ref{a2}).\footnote{%
Efficiency would only imply that a bad $k$ and a good $\ell $ cannot be both
consumed by both agents.} Fix a bad $k$ and a good $\ell $ s. t. $%
z_{1k},z_{1\ell }>0$. Buying $\frac{1}{|p_{k}|}$ unit of $k$ and $\frac{1}{%
p_{\ell }}$ unit of $\ell $ is budget neutral so by competitiveness it is
not profitable to agent $2$: $\frac{u_{2k}}{|p_{k}|}+\frac{u_{2\ell }}{%
p_{\ell }}\leq 0\Longleftrightarrow \frac{u_{2\ell }}{u_{2k}}\geq \frac{%
p_{\ell }}{p_{k}}$. But $z_{1}$ is $1$'s competitive demand so $\frac{%
u_{1\ell }}{p_{\ell }}=\frac{u_{1k}}{p_{k}}$: combining the last two
inequalities gives $\frac{u_{1\ell }}{u_{1k}}\leq \frac{u_{2\ell }}{u_{2k}}$
implying $k<\ell $ as claimed.

Together these three properties imply that at most one item $a$ can be
shared by both agents in the sense $z_{1a},z_{2a}>0$. Moreover if $G_{i}$
(resp. $B_{i}$) is the set of goods (resp. bads) consumed by agent $i$, then
all items in $B_{1}$ and $G_{2}$ are ranked below all items in $B_{2}$ and $%
G_{1}$, with at most one common item to both $B_{i}$-s or to both $G_{i}$%
-s.\smallskip

\noindent \textit{Step 3.2 We show} $|CU(\mathcal{P})|\leq 2m-1$

Consider a competitive allocation where no item is shared. By Step 1 there
is an index $k$ such that agent $2$ eats all goods in $\{1,\cdots ,k\}$ and
all bads in $\{k+1,\cdots ,m\}$, while agent $1$ eats the bads of $%
\{1,\cdots ,k\}$ and the goods of $\{k+1,\cdots ,m\}$. There are at most $%
m-1 $ such allocations.

Now consider a competitive allocation where our agents split (only) item $k$%
. The assignment of all other items to one agent or the other is determined
as in the previous paragraph. Thus the relative prices of all items eaten by
agent $i$ are determined by her marginal utilities, and the equality of both
budgets clinches the price and a single division of item $k$. Hence there
are at most $m$ competitive divisions splitting an item.\smallskip

\noindent \textit{Step 3.3 An example where} $|CU(\mathcal{P})|=2m-1$

There are only bads and the utilities generalize the example given just
after Proposition 1 in Subsection 6.1. We use the notation $(x)_{+}=\max
\{x,0\}$:%
\begin{equation*}
u_{1k}=-2^{(k-2)_{+}}\text{ for }1\leq k\leq m-1\text{ ; }u_{1m}=-2^{m-2}+1
\end{equation*}%
\begin{equation*}
u_{21}=-(2^{m-2}+1)\text{ ; }u_{2k}=-2^{(m-1-k)_{+}}\text{ for }2\leq k\leq m%
\text{ }
\end{equation*}%
We let the reader check that giving bads $1$ to $k$ to agent $1$ and the
rest to agent $2$ is a competitive allocation for budget $-1$ and the price:%
\begin{equation*}
p=-(\frac{u_{11}}{2^{k}},\cdots ,\frac{u_{1k}}{2^{k}},\frac{u_{2(k+1)}}{%
2^{m-k}},\cdots ,\frac{u_{2m}}{2^{m-k}})
\end{equation*}%
whereas splitting equally bad $k$, giving bads $1$ to $k-1$ to agent $1$,
and bads $k+1$ to $m$ to agent $2$, is a competitive allocation for budget $%
-1$ and the price:%
\begin{equation*}
p=-\frac{1}{3}(\frac{u_{11}}{2^{k-3}},\cdots ,\frac{u_{1(k-1)}}{2^{k-3}},%
\frac{u_{1k}}{2^{k-3}}=\frac{u_{2k}}{2^{m-2-k}},\frac{u_{2(k+1)}}{2^{m-2-k}}%
,\cdots ,\frac{u_{2m}}{2^{m-2-k}})
\end{equation*}

This example is \ clearly robust: small perturbations of the disutility
matrix preserve the number of competitive allocations.\smallskip

\noindent \textit{Step 3.4 }It remains to consider the case where the
sequence $\frac{u_{1k}}{u_{2k}}$ increase weakly but not strictly. Suppose $%
\frac{u_{1k}}{u_{2k}}=\frac{u_{1(k+1)}}{u_{2(k+1)}}$. Then in an competitive
allocation with price $p$ we have%
\begin{equation*}
\frac{u_{ik}}{p_{k}}=\frac{u_{i(k+1)}}{p_{k+1}}\text{ for }i=1,2
\end{equation*}%
Indeed if one of $i=1,2$ eats both $k$ and $k+1$, this follows by (\ref{3}).
If on the contrary $i$ eats item $k$ and $j$ eats item $k+1$, then (\ref{3})
again implies $\frac{u_{ik}}{p_{k}}\leq \frac{u_{i(k+1)}}{p_{k+1}}$ and $%
\frac{u_{j(k+1)}}{p_{k+1}}\leq \frac{u_{jk}}{p_{k}}$.

So for a given amount of money spent by $i$ on items $k$ and $k+1$, she gets
the same utility no matter how she splits this expense between the two items.%
\footnote{%
If one item is a good and the other a bad, shifting money between them
either increase both consumptions or decrease both.} Therefore there is an
interval of competitive allocations obtained by shifting the consumption of $%
k$ and $k+1$ while keeping the total expense on these two items fixed for
each agent. They all give the same utility profile and use the same price.
If we merge $k$ and $k+1$ with endowments $\omega _{k},\omega _{k+1}$ into
an item $k^{\ast }$ with one unit of endowment, $\omega _{k^{\ast }}=1$, and
utilities $u_{ik^{\ast }}=u_{ik}\omega _{k}+u_{i(k+1)}\omega _{k+1}$, the
above interval of competitive allocations becomes a single competitive
allocation for the new price $p_{k^{\ast }}=p_{k}\omega _{k}+p_{k+1}\omega
_{k+1}$, with $p$ unchanged elsewhere. By successively merging all the items
sharing the same ratio $\frac{u_{1k}}{u_{2k}}$, we do not change the number
of competitive allocations distinct utility-wise, and reach a problem with
fewer items where the ratios $\frac{u_{1k}}{u_{2k}}$ increase strictly in $k$%
. So we only need to prove the statement in this case.\smallskip

\noindent \textit{Step 4. Statement} $ii)\smallskip $

We fix a negative problem $\mathcal{P}=(N,A=\{a,b\},u)$. By our Theorem
there is at least one bad, i. e., $A_{-}$ is non empty. Suppose first that $%
b $ is a bad, while $a$ is a good: $a\in A_{+}$. As in the previous proof we
can assume $u_{ia}\geq 0$ for all $i$, with at least one strict inequality.
In a competitive allocation $z$ everyone consumes $b$ because all utilities
are negative. Some agent $i$ consumes $a$ as well, and by genericity and
Efficiency no other agent does. Moreover $i$ must have the highest ratio $%
\frac{u_{ia}}{u_{ib}}$: this determines the competitive price and it is then
easy to check that $CU(\mathcal{P})$ is unique, no matter how many agents
have the highest ratio $\frac{u_{ja}}{u_{jb}}$. We omit the details.

We turn the case where both items are bads. We label the agents $i\in
\{1,\cdots ,n\}$ in such a way that the ratios $\frac{u_{ia}}{u_{ib}}$
increase weakly in $i$. We describe first the efficient and non envious
allocations (which will be useful in the proof of Proposition 3), then the
competitive allocations in step 4.3.\smallskip

\noindent \textit{Step 4.1. Assume }$\frac{u_{ia}}{u_{ib}}$ \textit{%
increases strictly in }$i$. If $z$ is an efficient allocation, then for all $%
i,j$, $\{z_{ia}>0$ and $z_{jb}>0\}$ implies $i\leq j$. In particular at most
one agent is eating both bads, and we have two types of efficient and
envy-free allocations. For $1\leq i\leq n-1$ the $i/i+1$\textit{-cut} $%
z^{i/i+1}$ is the allocation $z_{j}^{i/i+1}=(\frac{1}{i},0)$ for $j\leq i$,
and $z_{j}^{i/i+1}=(0,\frac{1}{n-i})$ for $j\geq i+1$. For $2\leq i\leq n-1$
the allocation $z$ is an $i$\textit{-split} if there are numbers $x,y\ $such
that%
\begin{equation}
z_{j}=(\frac{1-x}{i-1},0)\text{ for }j\leq i-1\text{ ; }z_{j}=(0,\frac{1-y}{%
n-i})\text{ for }j\geq i+1  \label{8}
\end{equation}%
\begin{equation}
z_{i}=(x,y)\text{ with }0\leq x\leq \frac{1}{i}\text{, }0\leq y\leq \frac{1}{%
n-i+1}  \label{11}
\end{equation}%
Also, $z$ is a $1$-split if $z_{1}=(1,y)$ and $z_{j}=(0,\frac{1-y}{n-1})$
for $j\geq 2$; and $z$ is a $n$-split if $z_{n}=(x,1)$ and $z_{j}=(\frac{1-x%
}{n-1},0)$ for $j\leq n-1$. Note that the cut $z^{i/i+1}$ is both an $i$%
-split and an $i+1$-split.

We have shown that, if $\frac{u_{ia}}{u_{ib}}$ increases strictly, an
efficient and envy-free allocation must be an $i$-split. We turn to the case
where the increase is not strict.\smallskip

\noindent \textit{Step 4.2}. Assume the sequence $\frac{u_{ia}}{u_{ib}}$
increases only weakly, for instance $\frac{u_{ia}}{u_{ib}}=\frac{u_{(i+1)a}}{%
u_{(i+1)b}}$. Then if $z$ is efficient and envy-free we may have $%
z_{(i+1)a}>0$ and $z_{ib}>0$, however we can find $z^{\prime }$ delivering
the same utility profile and such that one of $z_{(i+1)a}^{\prime }$ and $%
z_{ib}^{\prime }$ is zero. Indeed No Envy and the fact that $u_{i}$ and $%
u_{i+1}$ are parallel gives $u_{i}\cdot z_{i}=u_{i}\cdot z_{i+1}$ and $%
u_{i+1}\cdot z_{i+1}=u_{i+1}\cdot z_{i}$, from which the claim follows
easily. We conclude that the $i$\textit{-}split allocations contain,
utility-wise, all efficient and envy-free allocations.\smallskip

\noindent \textit{Step 4.3.} If the cut $z^{i/i+1}$ is a competitive
allocation, the corresponding price is $p=-(i,n-i)$, and property (\ref{a2})
reads $\frac{u_{ja}}{i}\geq \frac{u_{jb}}{n-i}$ for $j\leq i$, $\frac{u_{jb}%
}{n-i}\geq \frac{u_{ja}}{i}$ for $j\geq i+1$, which boils down to%
\begin{equation}
\frac{u_{ia}}{u_{ib}}\leq \frac{i}{n-i}\leq \frac{u_{(i+1)a}}{u_{(i+1)b}}%
\text{ for }1\leq i\leq n-1  \label{9}
\end{equation}

Next for $2\leq i\leq n-1$ if the $i$\textit{-}split allocation $z$ (\ref{8}%
) is competitive, the (normalized) price must be $p=-n(\frac{u_{ia}}{%
u_{ia}+u_{ib}},\frac{u_{ib}}{u_{ia}+u_{ib}})$ and each agent must be
spending exactly $-1$:%
\begin{equation*}
p_{a}\frac{1-x}{i-1}=p_{b}\frac{1-y}{n-i}=p_{a}x+p_{b}y=-1
\end{equation*}%
which gives%
\begin{equation}
x=\frac{1}{nu_{ia}}{\large ((n-i+1)u}_{ia}-(i-1)u_{ib}{\large )}\text{ ; }y=%
\frac{1}{nu_{ib}}{\large (iu}_{ib}-(n-i)u_{ia}{\large )}  \label{50}
\end{equation}%
We let the reader check that these formulas are still valid when $i=1$ or $%
i=n-1$.

An $i$\textit{-}split allocation $z$ is \textit{strict }if it is not a cut,
i. e., both $x,y$ in (\ref{8}) are strictly positive. By (\ref{50}), for any 
$i\in \{1,\cdots ,n\}$ there is a strict $i$-split allocation that is
competitive if and only if

\begin{equation}
\frac{i-1}{n-i+1}<\frac{u_{ia}}{u_{ib}}<\frac{i}{n-i}  \label{10}
\end{equation}%
(with the convention $\frac{1}{0}=\infty $).\smallskip

\noindent \textit{Step 4.4}. Counting competitive allocations. There are at
most $n$ competitive (strict) $i$-split allocations, and $n-1$ cuts $%
z^{i/i+1}$, hence the upper bound $2n-1$. An example where the bound is
achieved uses any sequence $\frac{u_{ia}}{u_{ib}}$ meeting (\ref{10}) for
all $i\in \{1,\cdots ,n\}$, as these inequalities imply (\ref{9}) for all $%
i\in \{1,\cdots ,n-1\}$.

\subsection{Proposition 2}

We pick a division rule $f$ with associated welfare rule $F$ meeting ETE,
SOL and ILB, and we fix an arbitrary problem $\mathcal{P}=(u,\omega )$
(omitting $N,A$ that stay constant throughout the proof). We choose a
competitive division $(z,p,\beta )$ at $\mathcal{P}$ with associated utility
profile $\overline{U}$, and must show that $z\in f(\mathcal{P)}$.

\noindent Case 1: $\mathcal{P}$ is null. Then there is no feasible profile $%
U^{\prime }$ in $%
%TCIMACRO{\U{211d} }%
%BeginExpansion
\mathbb{R}
%EndExpansion
_{+}^{N}\diagdown \{0\}$ so SOL implies $F(\mathcal{P)\in }%
%TCIMACRO{\U{211d} }%
%BeginExpansion
\mathbb{R}
%EndExpansion
_{-}^{N}$. There the null utility if Pareto dominant, so $F(\mathcal{P)}%
=\{0\}$ by Efficiency.

\noindent Case 2: $\mathcal{P}$ is positive. Then $\beta =1$ and $p\cdot
z_{i}=1$ or $0$, respectively when $i$ is in $N_{+}$ or $N_{-}$. Consider
the positive problem $\mathcal{Q}=(w,\omega )$ where $w_{i}=p$ for $i\in
N_{+}$, and $w_{i}=0$ for $i\in N_{-}$. Efficiency implies that at least one
coordinate of $F(\mathcal{Q})$ is strictly positive, so by SOL they are all
non negative. Thus $F_{i}(\mathcal{Q})=0$ in $N_{-}$. By ETE $F_{i}(\mathcal{%
Q})$ does not depend on $i\in N_{+}$, moreover $W$ equal to $1$ in $N_{+}$
and $0$ in $N_{-}$ is Pareto optimal at $\mathcal{Q}$: we conclude that $F(%
\mathcal{Q})=W$.

Now we set $\overline{w}_{i}=\overline{U}_{i}p$ for all $i\in N$ (so $%
\overline{w}_{i}=w_{i}=0$ in $N_{-}$), and $\overline{\mathcal{P}}=(%
\overline{w},\omega )$. By the scale invariance property in Definition 2 we
have $F(\overline{\mathcal{P}})=\overline{U}$, moreover $\overline{w}%
_{i}\cdot z_{i}=\overline{U}_{i}(p\cdot z_{i})=\overline{U}_{i}$ in $N$. By
Pareto-Indifference (Definition 2) we conclude $z\in F(\overline{\mathcal{P}}%
)$.

We compare now $u$ and $\overline{w}$. Fix $a\in A_{+}\cup A_{-}$; for all $%
i\in N$ we claim%
\begin{equation*}
z_{ia}>0\Longrightarrow u_{ia}=\overline{U}_{i}p_{a}=\overline{w}_{i}\text{
; }z_{ia}=0\Longrightarrow u_{ia}\leq \overline{U}_{i}p_{a}=\overline{w}_{ia}
\end{equation*}%
Both claims are from (\ref{a2}) in Lemma 6 for $i\in N_{+}$; for $i\in N_{-}$
we must have $z_{ia}=0$ and we know $u_{ia}\leq 0$. The two statements
remain true for $a\in A_{0}$ because if $i$ eats some $a$ then $u_{ia}=0$,
and $p_{a}=0$ implies $\overline{w}_{ia}=0$ for all $i$.

Finally we apply ILB by lowering each $\overline{w}_{ia}$ to $u_{ia}$
whenever possible and $z\in f(\mathcal{P)}$ follows.

\noindent Case 3: $\mathcal{P}$ is negative. The omitted proof is similar,
only simpler because we do not need to distinguish between $N_{+}$ and $%
N_{-} $.\smallskip

We check finally that ILB is a consequence of Maskin Monotonicity (MM; see 
\cite{Ma}) in the additive domain. We do this in the case of bads only, as
both cases are similar. Individual allocations $z_{i}$ vary in the rectangle 
$[[0,\omega ]]$ ($0\leq z\leq \omega $)\ and utilities in $%
%TCIMACRO{\U{211d} }%
%BeginExpansion
\mathbb{R}
%EndExpansion
_{-}^{A}$, so MM for the division rule $f$\ means that for any two problems $%
\mathcal{P},\mathcal{P}^{\prime }$ on $N,A$\ and $z\in f(\mathcal{P})$ we
have%
\begin{equation}
\forall i\in N\{\forall w\in \lbrack \lbrack 0,\omega ]]\text{:\ }u_{i}\cdot
z_{i}>u_{i}\cdot w\Longrightarrow u_{i}^{\prime }\cdot z_{i}>u_{i}^{\prime
}\cdot w\}\Longrightarrow z\in f(\mathcal{P}^{\prime })  \label{45}
\end{equation}%
We fix $\mathcal{P},i\in N$ and $z\in f(\mathcal{P})$. We write $%
A^{0}=\{a|z_{ia}=0\},A^{1}=\{a|z_{ia}=\omega _{a}\}$\ and $A^{2}=A\diagdown
(A^{0}\cup A^{1})$. The implication in the premises of (\ref{45}) reads%
\begin{equation*}
\forall w\in \lbrack \lbrack 0,\omega ]]\text{ }u_{i}\cdot
(w-z_{i})<0\Longrightarrow u_{i}^{\prime }\cdot (w-z_{i})<0
\end{equation*}%
The cone generated by the vectors $w-z_{i}$ when $w$ covers $[[0,\omega ]]$
is $C=\{\delta \in 
%TCIMACRO{\U{211d} }%
%BeginExpansion
\mathbb{R}
%EndExpansion
^{A}|\delta _{a}\geq 0$ for $a\in A^{0}$, $\delta _{a}\leq 0$ for $a\in
A^{1}\}$. By Farkas Lemma the implication $\{\forall \delta \in C:$ $%
u_{i}\cdot \delta <0\Longrightarrow u_{i}^{\prime }\cdot \delta <0\}$ means
that, up to a positive scaling factor,%
\begin{equation*}
u_{ia}^{\prime }=u_{ia}\text{ on }A^{2}\text{ ; }u_{ia}^{\prime }\leq u_{ia}%
\text{ on }A^{0}\text{ ; }u_{ia}^{\prime }\geq u_{ia}\text{ on }A^{1}
\end{equation*}%
Thus MM says that after lowering a lost bid on item $a$, or increasing one
that gets the all of $a$, the initial allocation will remain in the selected
set. Now ILB only considers lowering a lost bid, so it is only
\textquotedblleft half\textquotedblright\ of MM. The competitive rule fails
the other half.

\subsection{Lemma 5}

\noindent \textit{Step 1 Only two bads}

\noindent We use the notation and results in Step 4 of the proof of
Proposition 1. Fix a problem $(N,\{a,b\},u)$ where the ratios $r_{i}=\frac{%
u_{ia}}{u_{ib}}$ increase strictly in $i\in \{1,\cdots ,n\}$ and write $%
S^{i} $ for the closed rectangle of $i$-split allocations (\ref{8}), (\ref%
{11}): we have $S^{i}\cap S^{i+1}=\{z^{i/i+1}\}$ for $i=1,\cdots ,n-1$, and $%
S^{i}\cap S^{j}=\varnothing $ if $i$ and $j$ are not adjacent. We saw that
envy-free and efficient allocations must be in the connected union $\mathcal{%
B}={\Large \cup }_{i=1}^{n}S^{i}$ of these rectangles. Writing $\mathcal{EF}$
for the set of envy-free allocations, we describe now the connected
components of $\mathcal{A}=\mathcal{B\cap EF}$. Clearly the set of
corresponding utility profiles has the same number of connected components.

We let the reader check that the cut $z^{i/i+1}$ is EF (envy-free) if and
only if it is competitive, i. e. inequalities (\ref{9}) hold, that we
rewrite as:%
\begin{equation}
r_{i}\leq \frac{i}{n-i}\leq r_{i+1}  \label{38}
\end{equation}%
If $z^{i/i+1}$ is EF then both $S^{i}\cap \mathcal{EF}$ and $S^{i+1}\cap 
\mathcal{EF}$ are in the same component of $\mathcal{A}$ as $z^{i/i+1}$,
because they are convex sets containing $z^{i/i+1}$. If both $z^{i-1/i}$ and 
$z^{i/i+1}$ are EF, so is the interval $[z^{i-1/i},z^{i/i+1}]$; then these
two cuts as well as $S^{i}\cap \mathcal{EF}$ are in the same component of $%
\mathcal{A}$. And if $z^{i/i+1}$ is EF but $z^{i-1/i}$ is not, then the
component of $\mathcal{A}$ containing $z^{i/i+1}$ is disjoint from any
component of $\mathcal{A}$ in ${\Large \cup }_{1}^{i-1}S^{j}$ (if any),
because $S^{i}\cap {\Large \cup }_{1}^{i-1}S^{j}=\{z^{i-1/i}\}$; a
symmetrical statement holds if $z^{i-1/i}$ is EF but $z^{i/i+1}$ is not.

Finally if $S^{i}\cap \mathcal{EF}\neq \varnothing $ while neither $%
z^{i-1/i} $ nor $z^{i/i+1}$ is in $\mathcal{EF}$, the convex set $S^{i}\cap 
\mathcal{EF}$ is a connected component of $\mathcal{A}$ because it is
disjoint from $S^{i-1}\cap \mathcal{EF}$ and $S^{i+1}\cap \mathcal{EF}$, and
all three sets are compact. In this case we speak of an interior component
of $\mathcal{A}$. We claim that $S^{i}$ contains an interior component if
and only if%
\begin{equation*}
\frac{i-1}{n-i+1}<r_{i-1}<r_{i}<r_{i+1}<\frac{i}{n-i}
\end{equation*}%
where for $i=1$ this reduces to the two right-hand inequalities, and for $%
i=n $ to the two left-hand ones. The claim is proven in the next Step.

Now consider a problem with the following configuration:%
\begin{equation*}
r_{1}<r_{2}<\frac{1}{n-1}<\frac{3}{n-3}<r_{3}<r_{4}<r_{5}<\frac{4}{n-4}<
\end{equation*}%
\begin{equation*}
<\frac{6}{n-6}<r_{6}<r_{7}<r_{8}<\frac{7}{n-7}<\frac{9}{n-9}\cdots
\end{equation*}%
By inequalities (\ref{38}) we have $z^{i/i+1}\in \mathcal{EF}$ for $i=3q-1,$
and $1\leq q\leq \lfloor \frac{n}{3}\rfloor $, and no two of those cuts are
adjacent so they belong to distinct components. Moreover $S^{i}$ contains an
interior component of $\mathcal{A}$ for $i=3q-2,$ and $1\leq q\leq \lfloor 
\frac{n+2}{3}\rfloor $, and only those. So the total number of components of 
$\mathcal{A}$ is $\lfloor \frac{n}{3}\rfloor +\lfloor \frac{n+2}{3}\rfloor
=\lfloor \frac{2n+1}{3}\rfloor $ as desired.

We let the reader check that we cannot reach a larger number of
components.\smallskip

\noindent \textit{Step 2: }$\{S^{i}$ \textit{contains an interior component}$%
\}\Longleftrightarrow \{$\textit{inequalities (\ref{12}) hold}$\}$

\noindent Pick $z\in S^{i}$ as in (\ref{8}), (\ref{11}) and note first that
for $2\leq i\leq n-1$, the envy-freeness inequalities reduce to just four
inequalities: agents $i-1$ and $i$ do not envy each other, and neither do
agents $i$ and $i+1$ (we omit the straightforward argument). Formally%
\begin{equation}
\frac{1}{r_{i+1}}(\frac{1}{n-i}-\frac{n-i+1}{n-i}y)\leq x\leq \frac{1}{r_{i}}%
(\frac{1}{n-i}-\frac{n-i+1}{n-i}y)  \label{13}
\end{equation}%
\begin{equation*}
r_{i-1}(\frac{1}{i-1}-\frac{i}{i-1}x)\leq y\leq r_{i}(\frac{1}{i-1}-\frac{i}{%
i-1}x)
\end{equation*}%
In the (non negative) space $(x,y)$ define the lines $\Delta (\lambda )$: $%
y=\lambda (\frac{1}{i-1}-\frac{i}{i-1}x)$ and $\Gamma (\mu )$: $x=\mu (\frac{%
1}{n-i}-\frac{n-i+1}{n-i}y)$. As shown on Figure 4{\Large \ }when $\lambda $
varies $\Delta (\lambda )$ pivots around $\delta =(\frac{1}{i},0)$,
corresponding to $z^{i/i+1}$, and similarly $\Gamma (\mu )$ pivots around $%
\gamma =(0,\frac{1}{n-i+1})$, corresponding to $z^{i-1/i}$. The above
inequalities say that $(x,y)$ is in the cone $\Delta ^{\ast }$ of points
below $\Delta (r_{i})$ and above $\Delta (r_{i-1})$, and also in the cone $%
\Gamma ^{\ast }$ below $\Gamma (\frac{1}{r_{i}})$ and above $\Gamma (\frac{1%
}{r_{i}+1})$. Thus $\delta \in \Gamma ^{\ast }$ if and only if $z^{i/i+1}$
is EF, and $\gamma \in \Delta ^{\ast }$ if and only if $z^{i-1/i}$ is EF. If
neither of these is true $\gamma $ is above or below $\Delta ^{\ast }$ on
the vertical axis and $\delta $ is to the left or to the right of $\Gamma
^{\ast }$ the horizontal axis. But if $\gamma $ is below $\Delta ^{\ast }$
while $\delta $ is right of $\Gamma ^{\ast }$, the two cones do not
intersect and $S^{i}\cap \mathcal{EF}=\varnothing $; ditto if $\gamma $ is
above $\Delta ^{\ast }$ while $\delta $ is left of $\Gamma ^{\ast }$ (see
Figures 4A,4B,4C). Moreover $\gamma $ above $\Delta ^{\ast }$ and $\delta $
right of $\Gamma ^{\ast }$ is impossible as it would imply%
\begin{equation*}
\frac{1}{n-i+1}>\frac{r_{i}}{i-1}\text{ and }\frac{1}{i}>\frac{1}{r_{i}(n-i)}
\end{equation*}%
a contradiction. We conclude that $\{S^{i}\cap \mathcal{EF}\neq \varnothing $
and $z^{i-1/i},z^{i/i+1}\notin \mathcal{EF}\}$ holds if and only if $\gamma $
is below $\Delta ^{\ast }$ and $\delta $ is to the left of $\Gamma ^{\ast }$%
, which is exactly the system (\ref{12}).

\begin{figure}
\centering
\vskip -0.5 cm
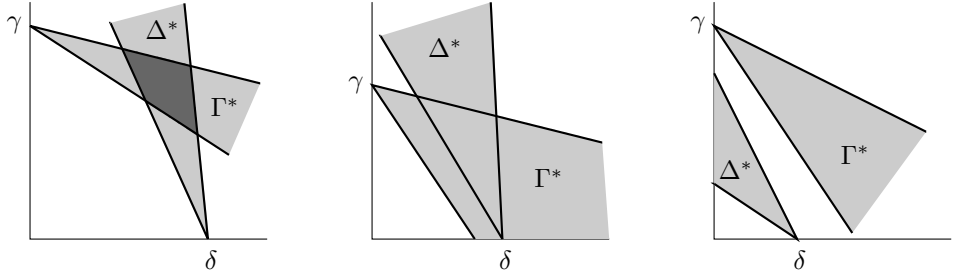
\caption*{Figures 4A, 4B, 4C}
\end{figure}

In the case $i=1$ the EF property of $z$ reduces to (\ref{13}) and the $i$%
-split allocation has $x=1$. If $r_{1}>\frac{1}{n-1}$ the right-hand
inequality in (\ref{13}) is impossible with $x=1$, therefore $r_{1}<\frac{1}{%
n-1}$; but then the fact that $z^{1/2}$ is not EF gives (see (\ref{38})) $%
r_{2}<\frac{1}{n-1}$ as desired. A similar argument applies for the case $%
i=n $.$\smallskip $

\noindent \textit{Step 3: Any number of bads}

\noindent Fix a problem $(N,\{a,b\},u)$ with $\lceil \frac{2n+1}{3}\rceil $
connected components as in Step 1. Given any $m\geq 3$, construct a problem $%
(N,\widetilde{A},\widetilde{u})$ with $\widetilde{A}=\{a,b_{1},\cdots
,b_{m-1}\}$ and for all agents $i$

\begin{equation*}
\widetilde{u}_{ia}=u_{ia}\text{ ; }\widetilde{u}_{ib_{k}}=\frac{1}{m-1}u_{ib}%
\text{ for all }1\leq k\leq m-1
\end{equation*}%
The bads $b_{k}$ are smaller size clones of $b$. If some $\widetilde{z}$ is
efficient and EF in the new problem, then the following allocation $z$ is
efficient and EF in the initial problem:%
\begin{equation*}
z_{ib}=\sum_{1}^{m-1}\widetilde{z}_{ib_{k}}\text{ ; }z_{ia}=\widetilde{z}%
_{ia}
\end{equation*}%
and $z,\widetilde{z}$ deliver the same disutility profile. Therefore in the
two problems the sets of efficient and EF allocations have the same number
of components.

\subsection{Proposition 3}

\noindent Fix a division rule $f$ meeting EFF and EVFR, and such that $F$ is
single-valued. In the problems discussed below, no two efficient and
envy-free allocations have the same utility profile, so $f$ is single valued
as well. Assume first $n=4$, $m=2$. Consider $\mathcal{P}^{1}$ where, with
the notation in the previous proof, we have%
\begin{equation*}
r_{1}<r_{2}<\frac{1}{3}<1<3<r_{3}<r_{4}
\end{equation*}%
(note that the numerical example at the beginning of Section 5 is of this
type)

By (\ref{38}) and (\ref{12}) $\mathcal{A}$ has three components: one
interior to $S^{1}$ (excluding the cut $z^{1/2}$), one around $z^{2/3}$
intersecting $S^{2}$ and $S^{3}$, and one interior to $S^{4}$ excluding $%
z^{3/4}$. Assume without loss that $f$ selects an allocation in the second
or third component just listed, and consider $\mathcal{P}^{2}$ where $%
r_{1},r_{2}$ are unchanged but the new ratios $r_{3}^{\prime },r_{4}^{\prime
}$ are%
\begin{equation*}
r_{1}<r_{2}<3<r_{3}^{\prime }<1<r_{4}^{\prime }<\frac{1}{3}
\end{equation*}%
Here, again by (\ref{38}) and (\ref{12}), $\mathcal{A}$ has a single
component interior to $S^{1}$, the same as in $\mathcal{P}^{1}$: none of the
cuts $z^{i/i+1}$ is in $\mathcal{A}$ anymore, and there is no component
interior to another $S^{i}$. When we decrease continuously $r_{3},r_{4}$ to $%
r_{3}^{\prime },r_{4}^{\prime }$, the allocation $z^{1/2}$ remains outside $%
\mathcal{A}$ and the component interior to $S^{1}$ does not move. Therefore
the allocation selected by $f$ cannot vary continuously in the ratios $r_{i}$%
, or in the underlying utility matrix $u$.

We can clearly construct a similar pair of problems to prove the statement
when $n\geq 5$ and $m=2$. And for the case $m\geq 3$ we use the cloning
technique in Step 3 of the previous proof.

\end{document}